\documentclass[aps,prd,amsmath,floats,floatfix,superscriptaddress,nofootinbib,showpacs,onecolumn]{revtex4}

\usepackage{amssymb}
\usepackage{amsmath}
\usepackage{verbatim}
\usepackage{mathrsfs}
\usepackage{amsfonts}
\usepackage{latexsym}
\usepackage{epsfig}
\usepackage{color}
\usepackage{graphicx,subfigure}
\usepackage{units}
\usepackage{overpic}
\usepackage{soul} 
\usepackage{graphicx}
\usepackage{physics}
\usepackage{slashed}
\usepackage[dvipsnames]{xcolor}
\usepackage{bm}
\usepackage{hyperref}
\usepackage[dvipsnames]{xcolor}
\hypersetup{
    colorlinks,
    linkcolor={red!50!black},
    citecolor={blue!50!black},
    urlcolor={blue!80!black}
}
\usepackage{orcidlink}

\begin{document}
\title{
Self-interacting scalar field distributions around Schwarzschild black holes 
}

\author{Alejandro Aguilar-Nieto\orcidlink{0000-0003-0859-0546}}
\affiliation{Instituto de Ciencias Nucleares, Universidad Nacional
  Aut\'onoma de M\'exico, Circuito Exterior C.U., A.P. 70-543,
  M\'exico D.F. 04510, M\'exico}

\author{V\'ictor Jaramillo\orcidlink{0000-0002-3235-4562}} 
\affiliation{Instituto de Ciencias Nucleares, Universidad Nacional
  Aut\'onoma de M\'exico, Circuito Exterior C.U., A.P. 70-543,
  M\'exico D.F. 04510, M\'exico}
  
\author{Juan Barranco\orcidlink{0000-0002-9511-6772}}
\affiliation{Departamento de F\'isica, Divisi\'on de Ciencias e Ingenier\'ias,
Campus Le\'on, Universidad de Guanajuato, Le\'on 37150, M\'exico}

\author{Argelia Bernal\orcidlink{0000-0003-3133-5957}}
\affiliation{Departamento de F\'isica, Divisi\'on de Ciencias e Ingenier\'ias,
Campus Le\'on, Universidad de Guanajuato, Le\'on 37150, M\'exico}

\author{Juan Carlos Degollado\orcidlink{0000-0002-8603-5209}} 
\affiliation{Instituto de Ciencias F\'isicas, Universidad Nacional Aut\'onoma 
de M\'exico, Apartado Postal 48-3, 62251, Cuernavaca, Morelos, M\'exico}

\author{Dar\'{\i}o N\'u\~nez\orcidlink{0000-0003-0295-0053}}
\affiliation{Instituto de Ciencias Nucleares, Universidad Nacional
  Aut\'onoma de M\'exico, Circuito Exterior C.U., A.P. 70-543,
  M\'exico D.F. 04510, M\'exico}

\date{\today}


\begin{abstract}

Long-lived configurations of massive scalar fields around black holes may form
if the coupling between the mass of the scalar field and the mass of the black hole is very small. 
In this work we analyze the effect of self-interaction in the distribution of the long-lived cloud surrounding a static black hole. We
consider both attractive and repulsive self-interactions.
By solving numerically the Klein Gordon equation on a fixed background in the frequency domain,
we find that the spatial distribution of quasi stationary states may be larger 
as compared to the non interacting case. We performed a time evolution to determine
the effect of the self-interaction on the life time of the configurations our findings indicate that the contribution of the self-interaction is subdominant.

\end{abstract}


                    
\maketitle

\section{Introduction}

Scalar fields are promising candidates 
to explain the nature of Dark Matter.
In a Scalar Field Dark Matter model it is assumed that a classical field makes up the main fraction of the dark matter of the Universe. The description relies on the fact that a coherently, weakly interacting, oscillating light scalar field
formed a Bose Einstein Condensate (BEC) in a non-relativistic and low-momentum state~\cite{Hu:2000ke,Matos:1999et,Matos:2000ng,Matos:2000ss,Sikivie:2009qn,Stadnik:2015, Stadnik:2016,Hui:2016ltb}.
The BEC has a phase space density that enables it to describe the density profile of the galactic dark matter halo for a convenient approximations.
Amongst the many proposed candidates,
Axions, particles introduced by Peccei and Quinn \cite{Peccei:1977hh}, are scalar fields with a non-zero vacuum expectation value and keep the CP invariance of the strong interactions in the Lagrangian
involving all Yukawa couplings. 
The effects of such scalar field and other axion-like particles (with a lower mass) in astrophysical and cosmological scenarios have been investigated extensively; see, e.g., Refs.~\cite{Arvanitaki:2009fg,Arvanitaki:2010sy,Sikivie:2014lha,Marsh:2013ywa,Porayko:2014rfa,Schive:2014dra,Marsh:2015wka,Marsh:2015xka,Matos:2003pe,Matos:2007zza}.

Since the detection of gravitational waves by the LIGO-Virgo-Kagra collaboration~\cite{LIGOScientific:2018mvr,LIGOScientific:2018jsj,LIGOScientific:2020ibl,LIGOScientific:2021djp}
and the recent developments in electromagnetic observations, in particular the image of the  
central black hole and its shadow of M87 and Sagittarius A* captured by the Event Horizon Telescope \cite{EventHorizonTelescope:2019dse,EventHorizonTelescope:2019ggy,EventHorizonTelescope:2022xnr},
there is a renewed interest in the physical processes that may occur in the vicinity of compact objects.

Massive bosonic fields may form quasi-bound states around a black holes ~\cite{ Damour:1976jd, Zouros:1979iw, Detweiler1980, Gaina_1992, Cardoso:2005vk, Lasenby:2002mc, Grain:2007gn}. In a Schwarzschild background, all of these states are unstable to decay, leaking part of the field towards the black hole~\cite{Dolan:2007mj,Witek:2012tr}.
These scalar configurations are characterized by instability timescales which are much longer than the timescale set by the mass of the central black hole.
Because the low rate of decay, the scalar field configurations may 
remain surrounding a black hole for large time scales
depending on the values of the parameters involved \cite{Barranco:2012qs}. 
Therefore, scalar fields around black holes represent  a very convenient set up to model the super massive black hole surrounded by a dark matter halo in galaxies. 
Quasi stationary solutions to the Klein-Gordon equation on a Schwarzschild background in that context, were described in detail by Barranco et al. in Refs.~\cite{Burt:2011pv,Barranco:2012qs,Barranco:2013rua}. The results obtained in those references, indicate that it is possible
that scalar field halos may last for cosmological time scales around super massive black holes. These long lived configurations composed by a complex, massive and non-self-interacting scalar field,  were found to be characterized essentially by two parameters, namely the integer $\ell$, associated with the angular distribution of the field and the dimensionless quantity formed by the mass coupling between the mass of the black hole $M$, and the mass of the scalar particle $m$,
$GM m/\hbar c$, which can also be interpreted as one half the ratio between the black hole horizon radius $r_h=2GM/c^2$
and the characteristic wavelength of the scalar field $\lambda_\phi=\hbar c/m$.

For a scalar particle, the simplest non gravitational interaction is a quartic self-interaction. This generalization can be achieved by expanding a potential about a symmetric minimum, and realizing that the quartic term is the most important interaction term for small amplitudes. From the point of view of a field theory the quartic self-interaction is the largest value in the exponent that allows a renormalizable theory.
In an astrophysical scenario, Colpi et al. \cite{Colpi86} show that self-interactions in boson stars can produce significant phenomenological changes. In
particular, they show that the upper limit on the
mass of a boson stars increases notably
compared to the noninteracting case. 
Such findings show that boson stars can have masses even larger than a solar mass. Furthermore, in some regimes self-interacting boson stars are compact enough to be considered black hole mimickers \cite{Liebling:2012fv,Lemos:2008cv,Guzman:2005bs,Mielke:2000mh,Torres2000,AmaroSeoane:2010qx}. 
From a cosmological perspective, the role of self-interaction is of great importance in dark matter models, particularly within complex scalar field models as in Refs. \cite{Li:2013nal, Suarez:2016eez} where it has been shown that with a repulsive self-interaction, the scalar field goes through a radiation-like stage in the Friedman evolution which in turn increases the effective number of relativistic degrees of freedom that matches the estimates of big bang nucleosynthesis (see also \cite{Gutierrez-Luna:2021tmq} for further discussion).

Quasi stationary states of attracting self-interacting scalar fields where found in the context of gravitational collapse of unstable boson stars in Ref. \cite{Escorihuela-Tomas:2017uac}.
The evolution of self-interacting boson stars in spherical symmetry 
was performed by solving the Einstein Klein Gordon system numerically. When describing an unstable configuration, boson stars collapsed forming a black hole surrounded by a remnant of scalar field leaving long lasting states around the newly formed black hole.

In this work, 
we solve the Klein-Gordon equation describing a self-interacting scalar field in the background of a Schwarzschild black hole. We assume the field oscillates coherently and analyse the system in the frequency domain to find resonant states. 
We analyze in detail the consequences of a quartic self-interaction in the distribution of the field and present a thorough analysis of the quasi stationary solutions.
We observe that the self-coupling have important consequences in the phenomenology associated with scalar field distributions. We further analyse the behaviour in time of the scalar distribution.

The paper is organized as follows: In Sec.\ref{sec:setup} we present the formulation, the background spacetime and method of construction of solutions.
In Sec. \ref{sec:res} 
we present some examples of solutions of the Klein Gordon equation and describe the properties of resonant modes.
In Sec. \ref{sec:timedomain} we discuss the time development of resonant modes.
Finally, we summarize our results and present concluding remarks in Sec.\ref{sec:conclusions}. Throughout the paper we use geometric units such that $c=1=G$.

\section{Set up}
\label{sec:setup}
In this work, we will focus on the regime where self-interaction effects on the field become significant before gravitational backreaction and therefore, we shall restrict to a fixed black hole spacetime.
We  consider a complex massive scalar field $\Phi$ with mass $m$ described by the action $S=-\int d^4 x\sqrt{-g}  [\nabla^\sigma \Phi^* \nabla_\sigma\Phi+V(|\Phi|)]$ with 

\begin{equation}
\begin{split}
V(|\Phi|)&=\mu^2|\Phi|^2+\frac{1}{2}\,\eta\,\lambda |\Phi|^4.
\end{split}
\end{equation}
where $\mu=m/\hbar$ and for convenience we consider $\lambda > 0$. The attractive or repulsive nature of the self-interaction is set by $\eta=1$ or $\eta=-1$ respectively.

The stress-energy tensor of the scalar field reads 
\begin{eqnarray}\label{eq:setensor}
T_{\alpha\beta} = \frac{1}{2}\left(\nabla_{\alpha} \Phi \nabla_{\beta} \Phi^* +\nabla_{\beta} \Phi \nabla_{\alpha} \Phi^* - g_{\alpha\beta} [\nabla^\sigma \Phi \nabla_\sigma \Phi^* + V(|\Phi|)]\right) \ .
\end{eqnarray}
The conservation of the stress energy tensor $\nabla_{\nu}T^{\mu\nu}=0$ provides the corresponding equation of motion for the field, the Klein-Gordon equation:
\begin{eqnarray}
\nabla_{\alpha}\nabla^{\alpha} \Phi=\frac{d V(|\Phi|)}{d\Phi^*} \ .
\end{eqnarray}
and its complex conjugate.
We consider the background space-time as the non-rotating Schwarzschild black hole in Boyer-Lindquist coordinates $(t,r,\theta,\varphi)$ with element of line given by
\begin{equation}
ds^2=-N(r) dt^2+N^{-1}(r) dr^2+r^2 d\Omega^2,
\end{equation}
 with $N(r)=1-2M/r$, and $d\Omega^2=d\theta^2+\sin^2\theta d\varphi^2$. 
We assume that the scalar field is spherically symmetric and thus can be written as
\begin{equation}\label{eq:decomposition}
    \Phi(t,r)=\frac{u(r)}{r} e^{i\omega t}.
\end{equation}
where $\omega$ is a real number.
The resulting Klein-Gordon equation takes the form
\begin{equation}\label{eq:KG}
    \left[-N(r)\frac{d}{d r}\left(N(r)\frac{d}{d r}\right)+N(r)\mathcal{U}(r)\right]u(r)=\omega^2 u(r) \ ,
\end{equation}
where, 
\begin{equation}\label{eq:kgsemipot}
    \mathcal{U}(r)=\frac{2M}{r^3}+\mu^2+\eta\,\lambda\,\frac{u^2(r)}{r^2}.
\end{equation}
We will use the fact that the static spacetime at hand has a timelike Killing vector to define a diagnostic quantity with respect to the stress energy of the system. 
The energy density of the scalar field measured by a static observer, $\rho=-{T_0}^0$, provides a useful measure of the distribution of the field in the spacetime, explicitly is given as 
\begin{equation}
    \rho=\frac{1}{2r^2}\left(\frac{\omega^2u^2}{N}+N\left(\frac{d u}{d r}\right)^2+\mathcal{U}(r)u^2 - \frac{d}{dr}\frac{N u^2}{r}\right).
    \label{Density}
\end{equation}
This quantity will be used to characterize the scalar field distribution in the following sections.
Another quantity of interest is the 
total mass-energy storaged by the scalar field obtained by the integration of the energy density
\begin{equation} \label{eq:mass_sf}
    M_\Phi=4\pi\int_{2M}^\infty \rho r^2 dr \ .
\end{equation}
This quantity gives an estimation of the amount of scalar field in the spacetime that can be compared with the mass of the black hole.
\subsection{Scaling properties}
\label{sec:scaling}
%
In order to explore the space of solutions of
the Klein Gordon equation \eqref{eq:KG}, we
define a new function $v(r)$ as
\begin{equation}
    v(r)=\sqrt{ \lambda}\ u(r) \ ,
\end{equation}
such that the function $\mathcal{U}(r)$ becomes
\begin{equation}
    \mathcal{U}(r)=\frac{2M}{r^3}+\mu^2+\eta\,\frac{v^2(r)}{r^2} \ .
\end{equation}
The function $v$ encodes the self-interaction parameter and allow us to explore the solution space of Eq. \eqref{eq:KG} providing an infinite set of solutions for each value of $\lambda$. 
Additionally, in order to better resolve the behaviour of the field in the region close to the horizon, we use 
the tortoise coordinate, 
$r*=r+2M\ln(r/2M-1)$, to rewrite Eq.~(\ref{eq:KG}) as
\begin{equation}\label{eq:KGv}
\begin{split}
    -\frac{d^2 v}{d {r^*}^2}+N(r) \mathcal{U}(r) v=\omega^2 v  \ ,
\end{split}
\end{equation}
where $r$ is given implicitly in the definition of  $r*$. One can thus eliminate $\lambda$ from the numerical task because always enters as a scale. 
The equation \eqref{eq:KGv}, posses a further rescaling property, 
which is specified in Table \ref{rescaling}.
This rescaling allows us to use dimensionless quantities 
such as $r/M$, $M\mu$, $M \omega$, etc.
to specify any quantity in terms of the mass of the black hole. The dependence on the parameters in the Klein Gordon equation is in  $M\mu$,
and the value of $\eta$ (that is $+1$ or $-1$). 
\begin{table}[h!]\centering
  \begin{tabular}{r c l}
    $M$ & $\mapsto$            & $\alpha\, M$  \\
    \hline\hline
    $r$ & $\mapsto$ & $\alpha r$\\
    $(\mu,\omega)$ & $\mapsto$ & $\alpha^{-1} (\mu,\omega)$\\
    $(u,\lambda)$ & $\mapsto$ & $(u,\lambda)$\\
    \hline \hline
  \end{tabular}\qquad
  \caption{Rescaling properties of the Klein-Gordon equation in the Schwazschild background Eq.~(\ref{eq:KG})
    \label{rescaling}}
\end{table}
%
\subsection{Asymptotically decaying solutions}

Equation \eqref{eq:KGv} represent a non-linear eingenvalue problem for the function $v$ and eigenvalue $\omega$.
Before presenting its solutions and the methods we employed to solve it, let us first emphasize some of the properties of the solution.
In the near horizon region $r^*\to-\infty$, the function $N(r)\to 0$, consequently solutions of Eq.~\eqref{eq:KGv}  may have the form
\begin{equation} \label{eq:close_bhv}
    v(r^*)\approx A\cos(\omega r^*-\delta) \ ,
\end{equation}
where the amplitude $A$ and the phase $\delta$ are real numbers. 
Notice however, that the function \eqref{eq:close_bhv} together with the ansatz \eqref{eq:decomposition}, contains no physical solutions since it represents both out-coming and in-going modes.
The physical situation we shall consider, a field distribution around a black hole, excludes outgoing modes.
In the following section we will address this issue in more detail.

Let us consider values of $\omega$ such that $\omega<\mu$. Then, in the limit $r^*\rightarrow\infty$
Eq.~(\ref{eq:KGv}) admits exponentially decaying or growing solutions of the form $v(r^*) \approx C\exp(-k r^*)+D\exp(+k r^*)$
with $k=\sqrt{\mu^2-\omega^2}$.
Since we are interested in describing localized solutions, consistent with the asymptotically flat metric,
we will focus on the
solutions with exponential decay and set $D=0$.

The previous descriptions concerns only the asymptotic behaviour of the solutions of Eq.~\eqref{eq:KGv}. With only this information, the possible values that $\omega$ can take are infinite and the spectra is continuous. 
In the next section we shall show that there is a set of discrete values of $\omega$ that allows solutions in which the in-going modes have a larger amplitude than the out-going modes. These solutions are thus closer to the physical description of having an event horizon as the left boundary.  

\begin{figure}
\begin{center}
\begin{center}
\includegraphics[width=0.5\textwidth, angle=0]{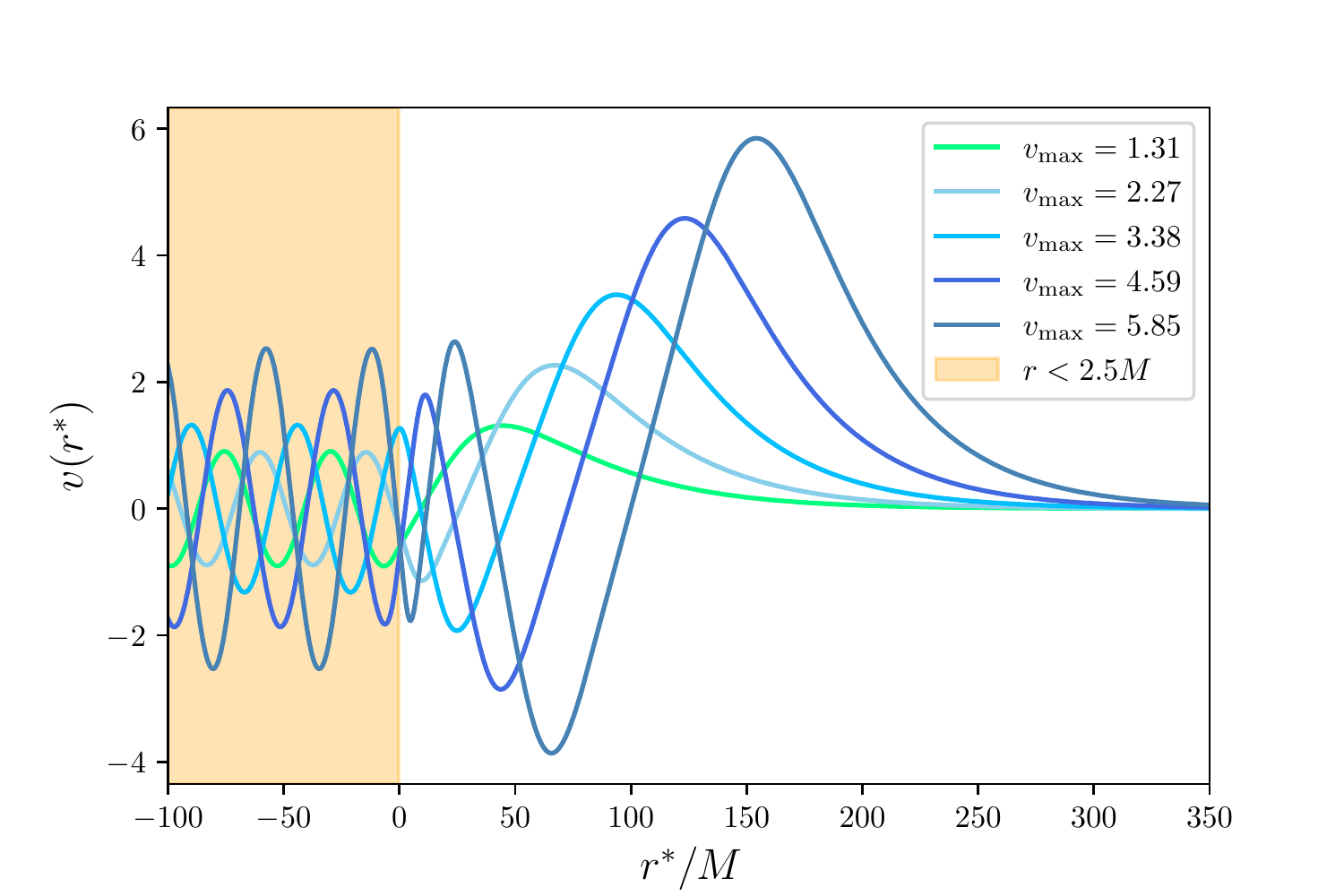} 
\end{center}
\caption{Sample of solutions of Eq.~\eqref{eq:KGv} with an exponential decay at infinity.
In the region close the horizon (shaded area,  $r*\rightarrow -\infty$) solutions behave according to Eq.~\eqref{eq:close_bhv}, the maximal amplitude in this region sets the value of $A$.
The value in the far region set by the $C$ also fixes the value of $v_{\max}$.
Thus, each solution can be characterized by the amplitudes $A$ and $v_{\max}$.
}
\label{fig:v_asintot}
\end{center}
\end{figure}

In order to find the solutions of 
Eq.~(\ref{eq:KGv}) we integrate it numerically as follows.
We fix $\eta$, $\mu$ and choose some value $\omega$.
Then we set a value of $v$ at the outermost point of the numerical grid 
${r^*}_{\rm max}$ and integrate the equation inwards up to the left boundary ${r^*}_{\rm min}$.
Fig.~\ref{fig:v_asintot} shows the plot of $v$ for $M\mu=0.14$ and $M\omega=0.1370$ obtained by this procedure. There is a region in which the behaviour \eqref{eq:close_bhv} is reached and the function $v$ is completely characterized by its maximum amplitude in that region $A$, and the asymptotic behaviour. 
Furthermore, we have found that for each solution $v(r)$, the quantity $v_\mathrm{max}$ defined as the local maximum attained by the function $v$ in the far horizon region, can be used to characterize that solution.
In the next section we will show that
comparing the amplitude in the far region, $v_{\rm max}$, with the amplitude near the horizon, $A$, it is possible to find a discrete set of values of $\omega$ that allows quasi bound or resonant configurations.  

\begin{figure}
\begin{center}
\begin{center}
\includegraphics[width=0.4\textwidth, angle=0]{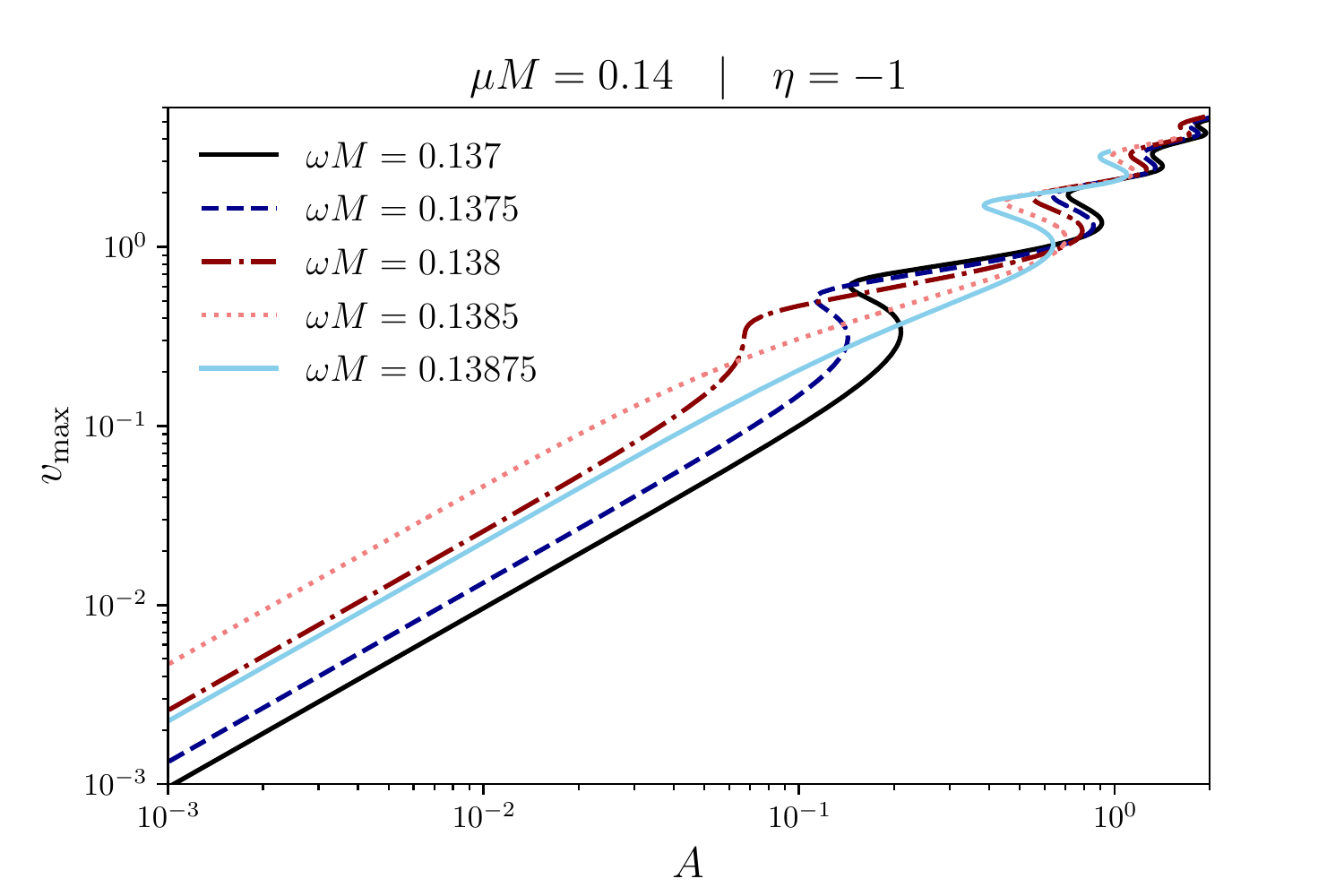} \includegraphics[width=0.4\textwidth, angle=0]{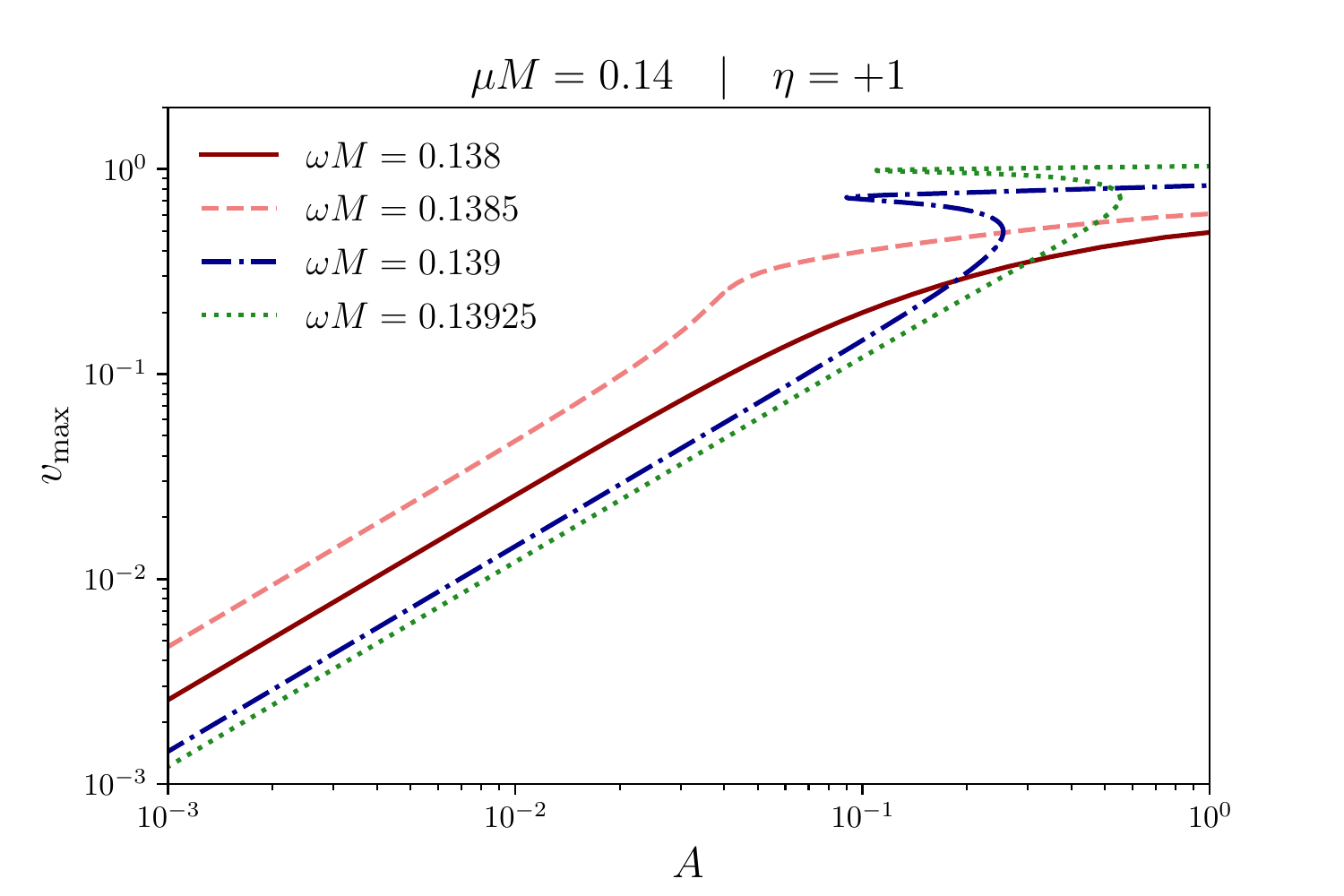}
\end{center}
\caption{
Amplitude near the horizon $A$, and $v_{\max}$, for solutions of Eq.~\eqref{eq:KGv}.
Each point of a curve characterize an
asymptotically decaying solution with a behaviour close the horizon given by Eq.~\eqref{eq:close_bhv}.
Without specifying the boundary conditions the spectra in $\omega$ is continuum. 
The weak self-interacting regime corresponds to the extreme left in both panels, where the behaviour is almost linear.
}
\label{fig:Avsdelta0}
\end{center}
\end{figure}

Decaying solution are characterized by the amplitudes $A$ and $v_{\rm max}$ for a given value of $\omega$.
Figs.~\ref{fig:Avsdelta0} display sets of solutions with the desired asymptotic behaviour for some values of $\omega$.
We will refer solutions with $v_{\rm max}\to0$ as solutions in the \textit{weak self-interacting regime}. As the value of $v_{\rm max}$ increases, the term $v^2$ in Eq.~(\ref{eq:KGv}) becomes important and defines solutions in the \textit{strong self-interacting regime}.
These functions however, are not physical in the sense that we have not yet imposed the boundary condition at the left. That is, 
these functions contain outgoing modes coming from the left which is not physically possible.

\section{Resonant states}
\label{sec:res}

\begin{figure}
\begin{center}
\begin{center}
\includegraphics[width=0.4\textwidth, angle=0]{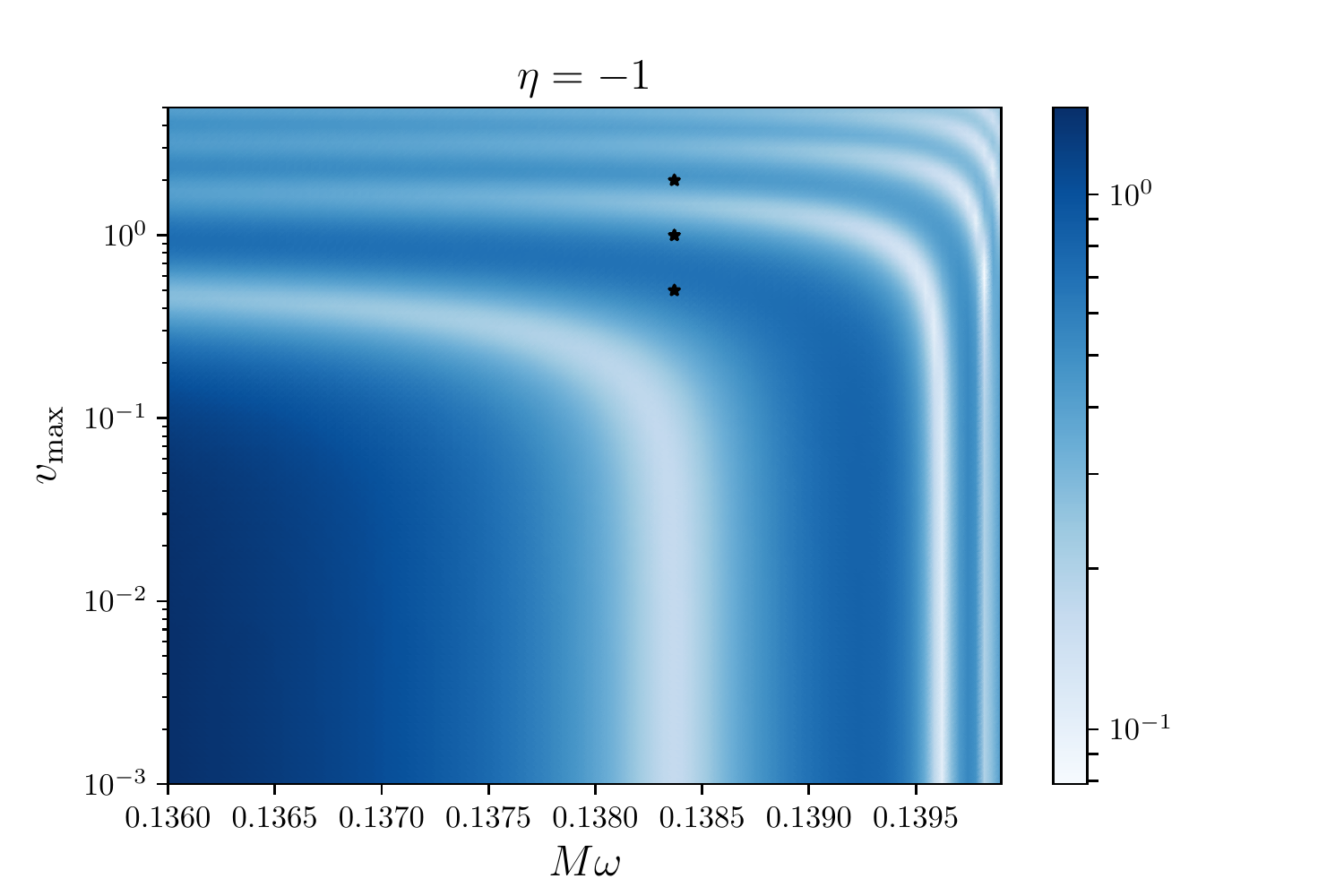}
\includegraphics[width=0.4\textwidth, angle=0]{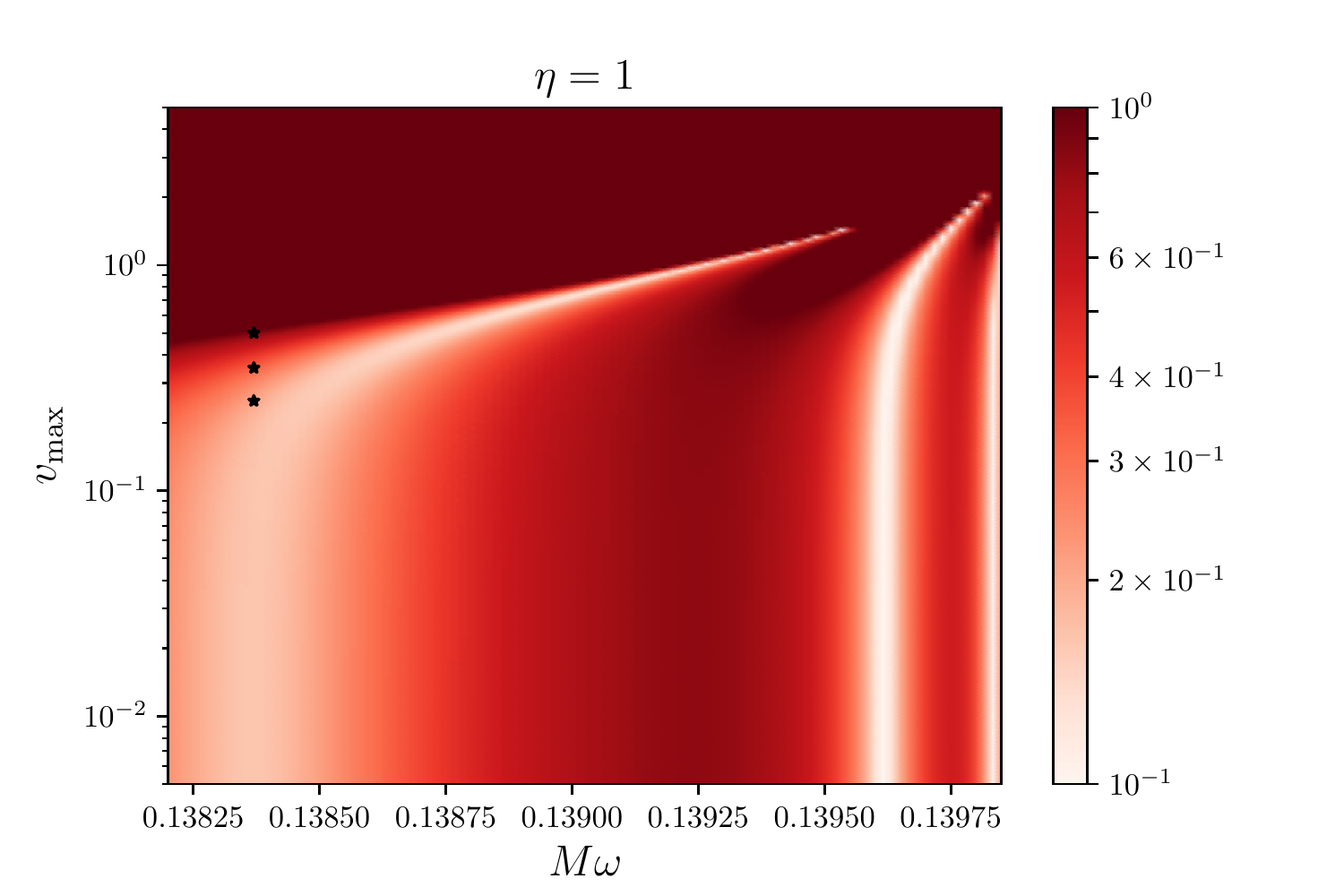}
\end{center}
\caption{Color map of the ratio $A/v_{\mathrm{max}}$ 
for $M\mu=0.14$. 
The intensity of the color indicates the value of the ratio.
The white bands, in which the value of $v_{\rm max}$ is much larger than $A$, correspond to resonant states. Left panel is for $\eta=-1$ and right panel is for $\eta=+1$. 
The weakly self-interacting regime correspond to the lower part in which $v_{\rm max}\rightarrow 0$. 
For larger values of $v_{\rm max}$ solutions can not be found for $\eta=+1$.
The asterisk symbols are the configurations of Figs.~\ref{fig:Soluciones negativas} and \ref{fig:Soluciones positivas} with frequency $\omega=0.13837$.
}
\label{fig:picos0}
\end{center}
\end{figure}

\begin{figure}
\begin{center}
\includegraphics[width=0.5\textwidth, angle=0]{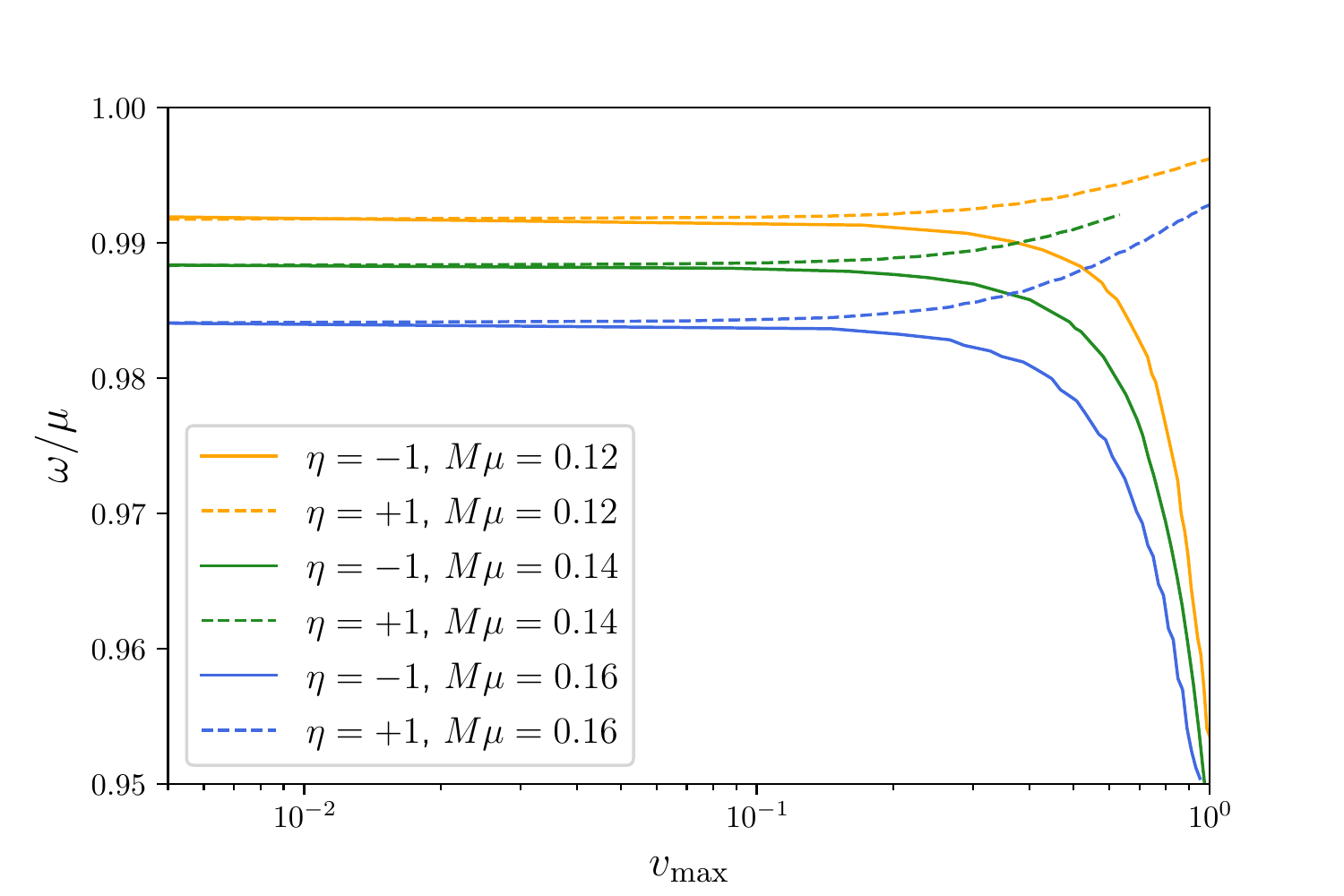}\\
\caption{Fundamental frequencies of resonant states, as a function of $v_{\rm max}$ for several cases of $M\mu$ for both, positive and negative self-interaction. For $\eta = +1$ (dashed line), as $v_{\rm max}$ grows, there are no resonant states.}
\label{fig:0p14}
\end{center}
\end{figure}

It is possible to quantify the amount of the scalar field that accumulates in the outer region with respect to the amplitude near the horizon by comparing the values of $v_{\rm max}$ and $A$. In Fig.~\ref{fig:picos0} we show a projection map of the ratio $A/v_{\mathrm{max}}$ using several values of $\omega M$ for $\mu M=0.14$. The left panel corresponds to $\eta=-1$ and the right panel corresponds to $\eta=1$. 
\begin{figure}
\begin{center}
\begin{center}
\includegraphics[width=0.4\textwidth, angle=0]{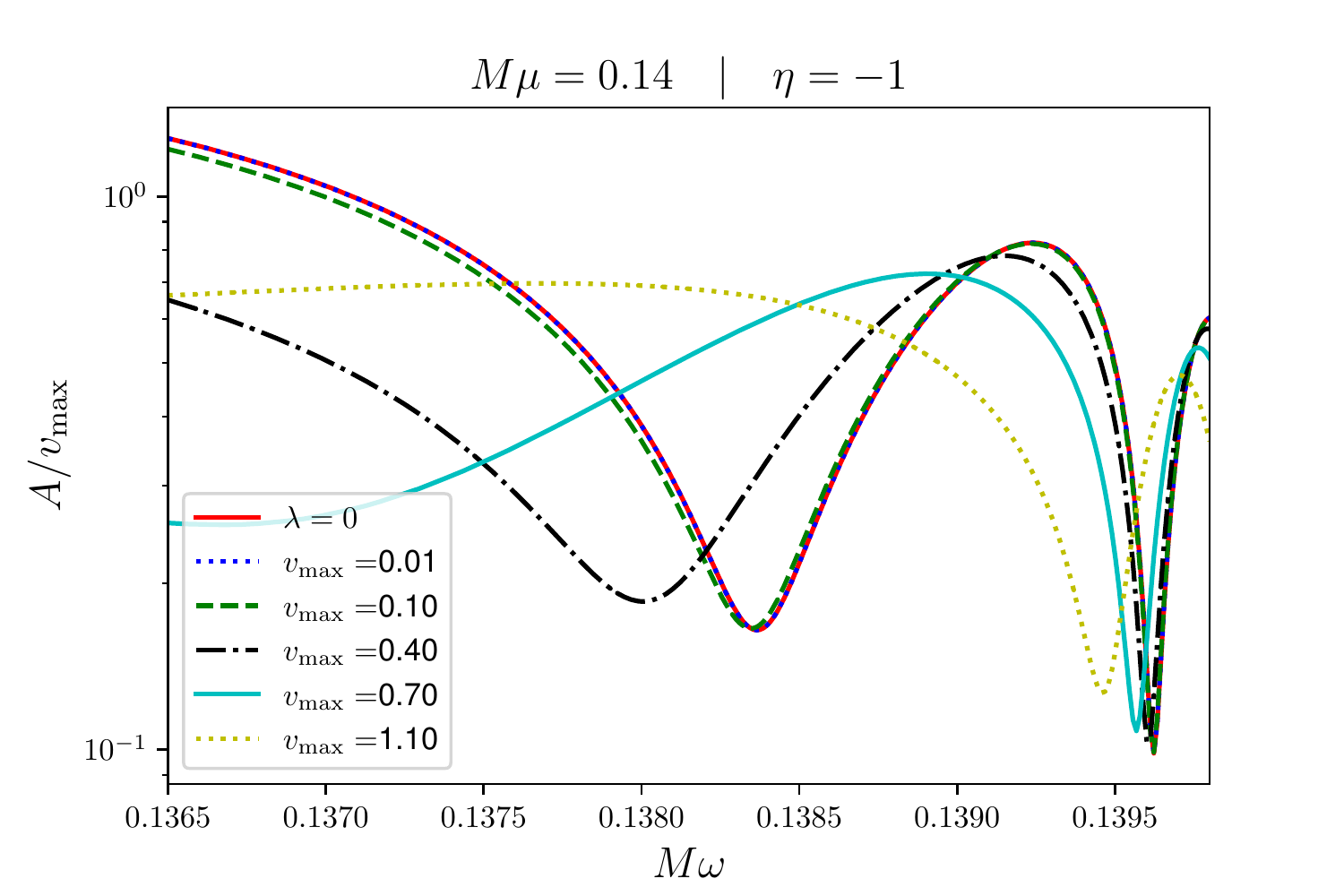}
\includegraphics[width=0.4\textwidth, angle=0]{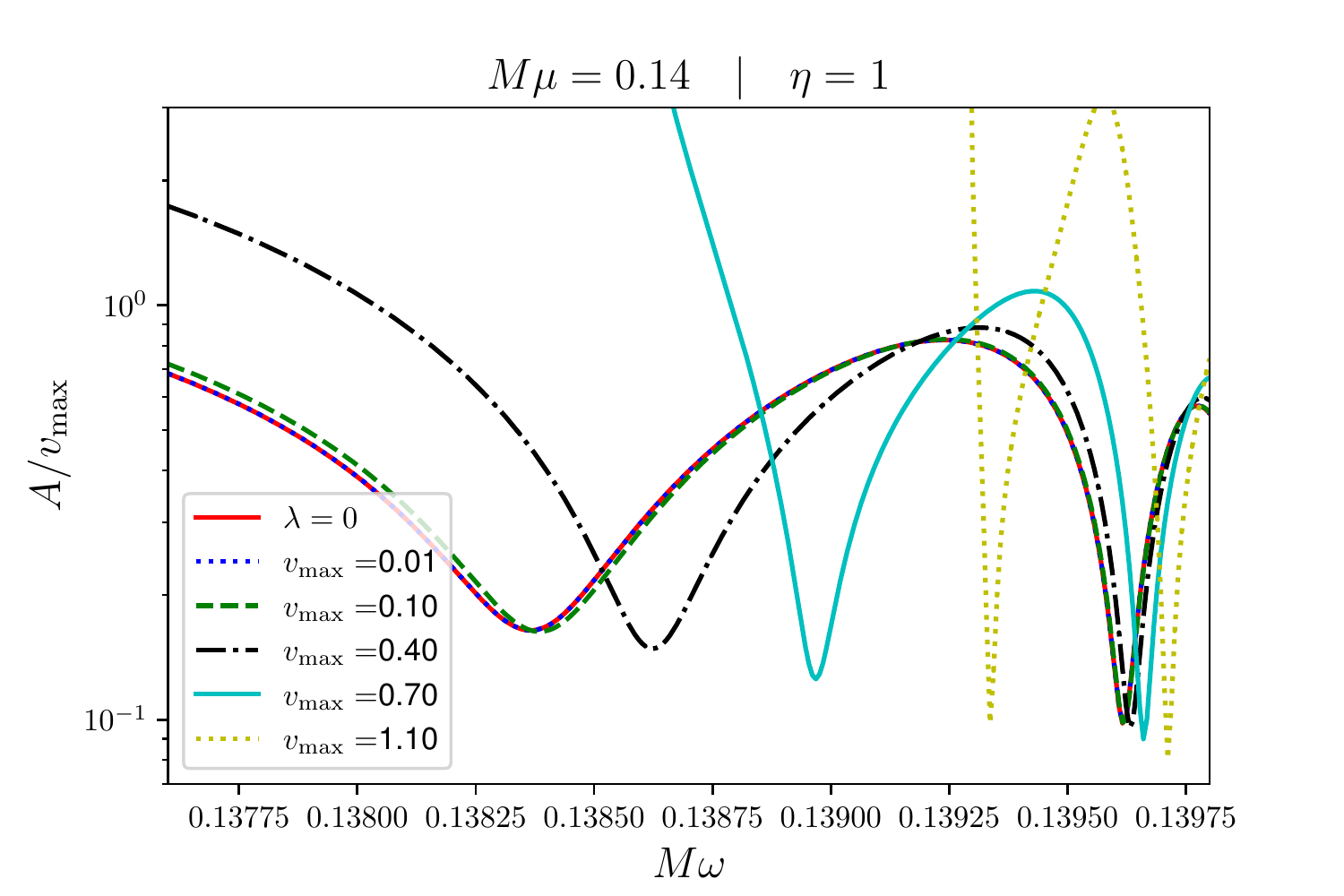}
\end{center}
\caption{Ratio A/$v_{\mathrm{max}}$ as a function of $\omega$ for different values of $v_{\mathrm{max}}$ and $M\mu=0.14$  
The local minima on each curve denote resonant solutions.
The case with $\lambda=0$ is also shown as a reference. In the weak regime the curves are almost indistinguishable from the non interacting case.}
\label{fig:cortes}
\end{center}
\end{figure}
In order to better visualize the frequencies of the resonant states and get some understanding on their change in $v_{\rm max}$, Fig,~\ref{fig:cortes} shows constant $v_\mathrm{max}$ cuts of the projected surface Fig.~\ref{fig:picos0}.
In these plots, resonant states are those solutions (represented by each curve) whose ratio $A/v_{\mathrm{max}}$ has a minimum. The value of the frequency at which the first minimum occurs corresponds to the frequency of the first resonant state (higher frequencies have been called overtones in close analogy to harmonic frequencies).
Consequently, configurations surrounding a static black hole with a larger amplitude far from the event horizon are characterized by a discrete set of frequencies $\omega$. This modes for $\lambda=0$ have been already characterized in Ref.~\cite{Barranco:2012qs}.
The effect of the self-interaction in the resonant states is to induce a change in the values of the frequencies as can be seen in Figs.~\ref{fig:picos0}.
The frequency of resonant states in the limit $v_{\rm max}$ through the vertical bands in Figs.~\ref{fig:picos0}
match the value of the fundamental and first resonant frequencies 
$M\omega = 0.13837 $ and 
$M\omega = 0.13956 $ respectively for  $M\mu=0.14$ and $\lambda=0$.
For $\eta=-1$, as $v_{\rm max}$ increases, the frequencies of the resonant states decrease with respect to the noninteracting case, while for $\eta=+1$ the frequencies increase.
Fundamental frequencies of resonant states are shown in Fig~\ref{fig:0p14}
as a function of $v_{\mathrm{max}}$ for some representative values of $M\mu$.
We have found that 
for $\eta=+1$ 
there might be no resonant states with higher values of $v_{\mathrm{max}}$. 
Other results related to the scalar field distribution as well as the issue about the lifetime of these states will be addressed below.

\subsection{Properties of resonant states}
\label{sec:analysis}
\begin{figure}
\begin{center}
\includegraphics[width=0.45\textwidth, angle=0]{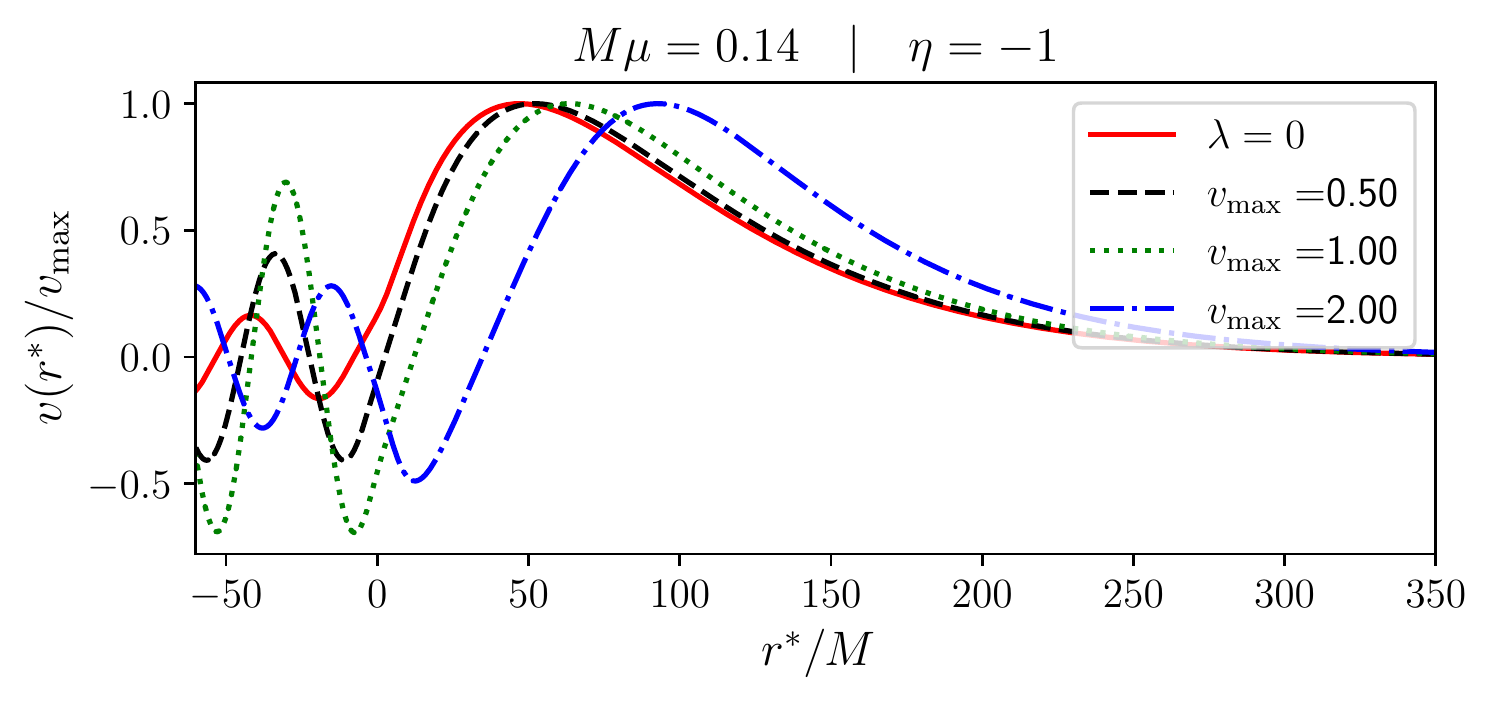}
\includegraphics[width=0.45\textwidth, angle=0]{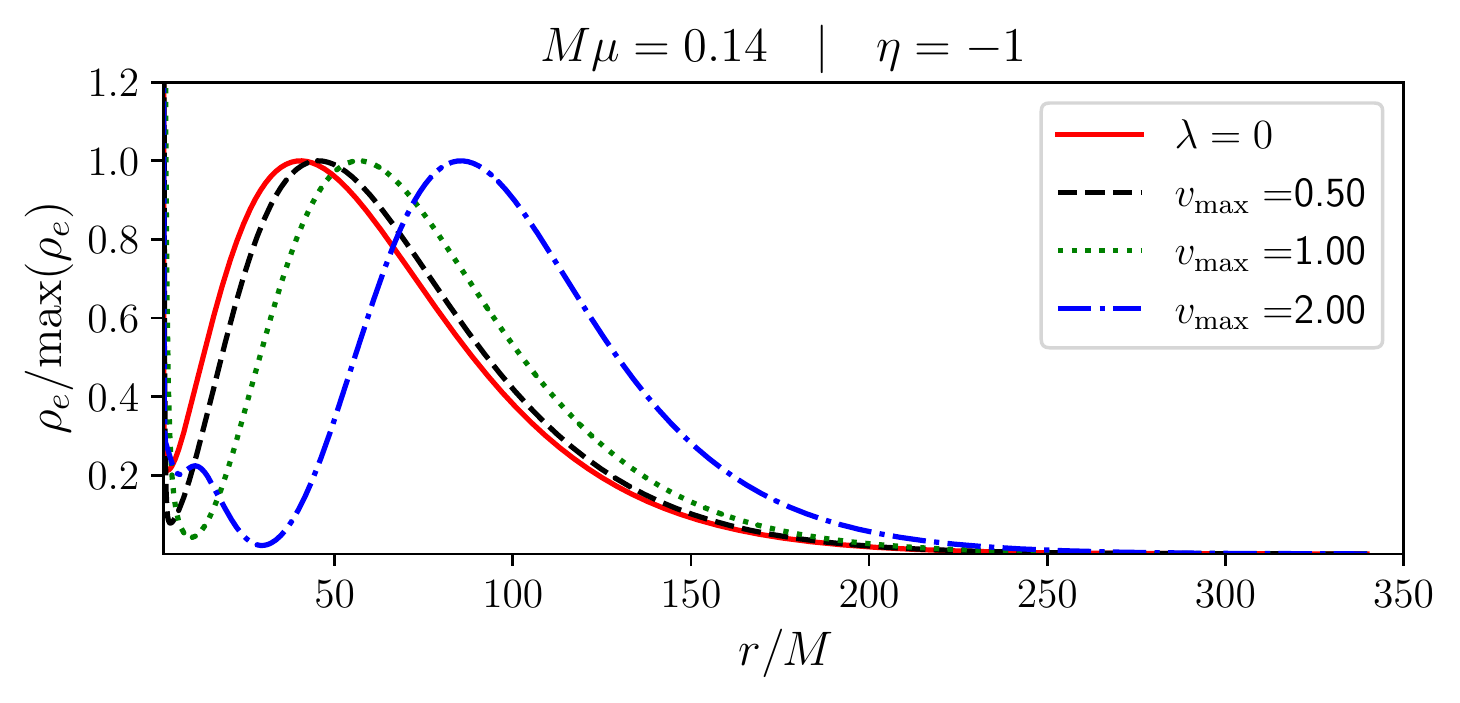}
\caption{The left panel shows the profile of the scalar field and the right panel the corresponding radial energy density for a fixed value of $M\omega =0.13837$ with different values of $v_{\mathrm{max}}$. The case  $\lambda=0$ is included for comparison. Both sets have been normalized for a better visualization.}
\label{fig:Soluciones negativas}
\end{center}
\end{figure}

\begin{figure}
\begin{center}
\includegraphics[width=0.45\textwidth, angle=0]{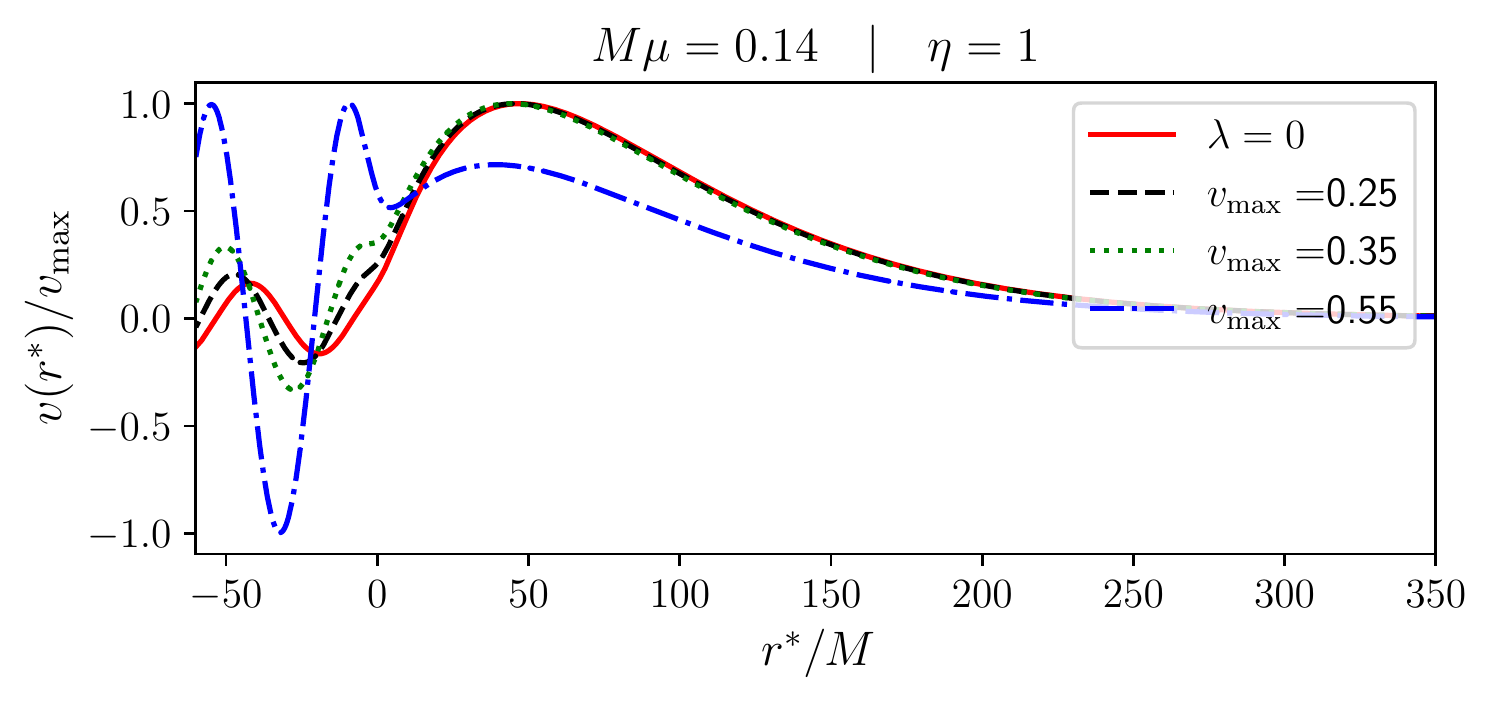}
\includegraphics[width=0.45\textwidth, angle=0]{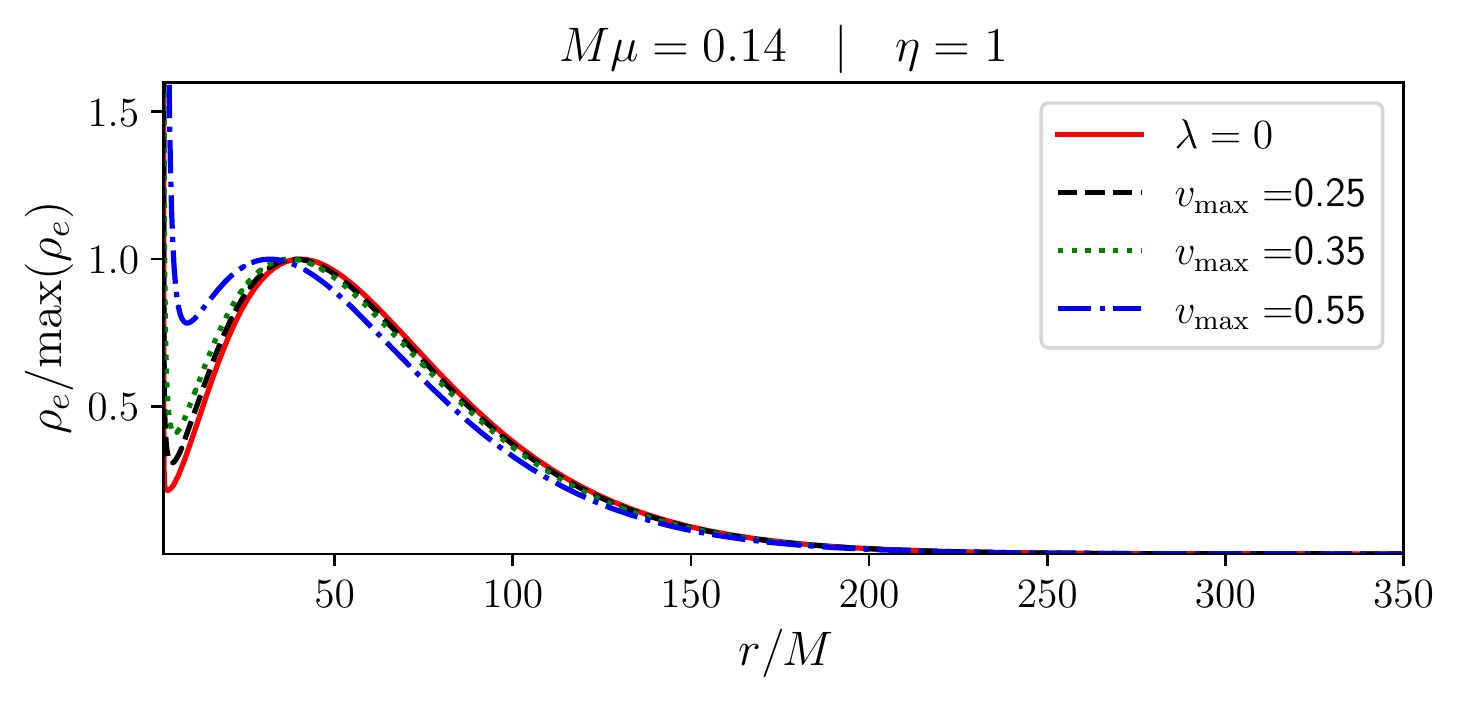}
\caption{Same as Fig. \ref{fig:Soluciones negativas} with $\eta=1$.}
\label{fig:Soluciones positivas}
\end{center}
\end{figure}

%
As stated before, the effect of the magnitude of $\lambda$ on the radial profile of the scalar field can be described in terms of $v_{\mathrm{max}}$. 
The first scenario we examine is the one where
$\eta=-1$ and $\omega$ is fixed. 
As a particular example, by taking the frequency $M\,\omega=0.13837$ (that corresponds to the first resonant state with $\lambda=0$),
configurations in the weak self-interacting regime  
are almost indistinguishable from the case with $\lambda=0$. 
In the strong regime the profiles differ slightly
because these configurations are no longer resonant states.
Fig.~\ref{fig:Soluciones negativas} displays the profiles $v(r^*)$ and 
the radial energy density $\rho_e=4\pi r^2\rho$, with $\eta=-1$ and $M\mu=0.14$ for some representative values of $v_{\mathrm{max}}$ in both weak and strong self-interacting regimes, the case 
with $\lambda=0$ is also shown for comparison.
The configurations in Fig.~\ref{fig:Soluciones negativas} correspond to the asterisk symbols in Fig.~\ref{fig:picos0}. For large values of $v_{\mathrm{max}}$, the ratio $A/v_{\rm max}$ becomes larger and it may happen that for large enough values of $v_{\mathrm{max}}$ along the fixed frequency, a profile with one or more nodes is found. 
Like in the $\lambda=0$ case, overtones correspond to solutions with increasing number of nodes as $\omega$ approaches $\mu$. 
In the negative self-interaction case,
$\eta=-1$, the effect of increasing $\lambda$ in the mass density is that the scalar field distribution spreads over a larger region and the maximum moves away from the horizon. 
The normalized mass density is shown in the right panel of 
Fig.~\ref{fig:Soluciones negativas}.
Given the behaviour \eqref{eq:close_bhv},
solutions of Eq. \eqref{eq:KGv} 
produce an infinite energy density at the horizon.
A positive self-interaction
$\eta=+1$ produces analogous changes in the radial profile of the scalar field, in this case however, the field concentrates closer the horizon and the changes are smaller in magnitude.
Another important difference is that for large values of $v_{\rm max}$ resonant solutions cease to exist.
Fig.~\ref{fig:Soluciones positivas} displays the radial scalar field profile and the radial energy density for $\eta=+1$ 

\begin{figure}
\begin{center}
\includegraphics[width=0.45\textwidth, angle=0]{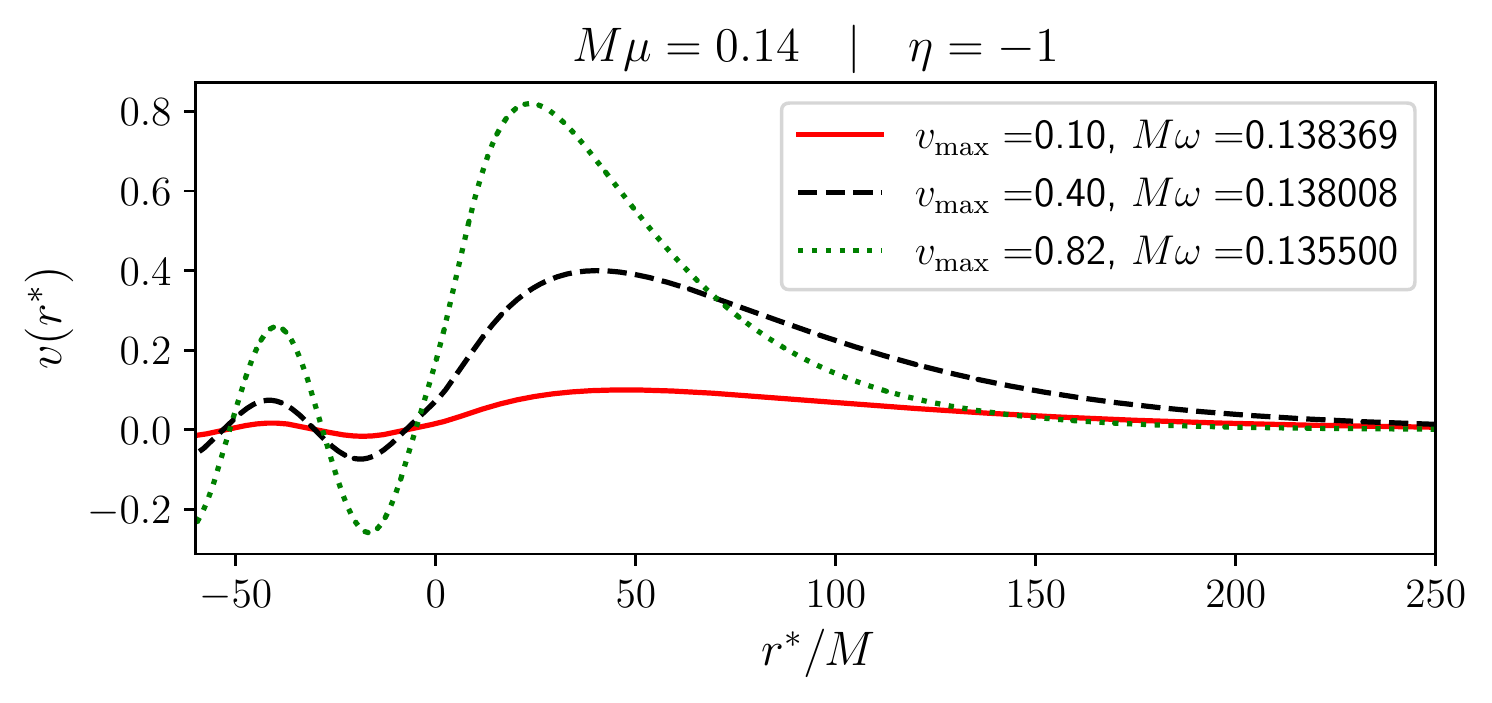}
\includegraphics[width=0.45\textwidth, angle=0]{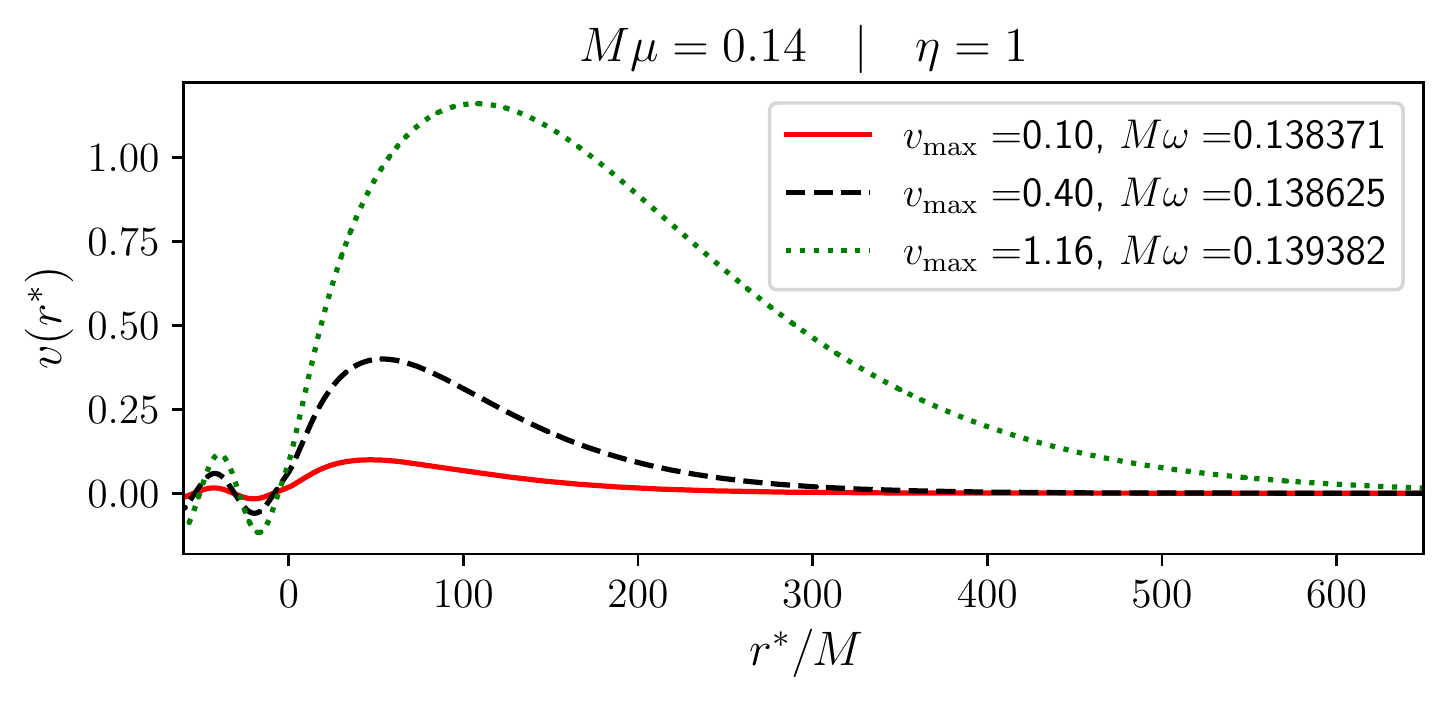}
\caption{Scalar field profile for the first resonant state for different values of $v_{\mathrm{max}}$ for $\eta=-1$, left panel, and  
$\eta=+1$, right panel.}
\label{fig:profiles_diff_eta}
\end{center}
\end{figure}

\begin{figure}
\begin{center}
\includegraphics[width=0.45\textwidth, angle=0]{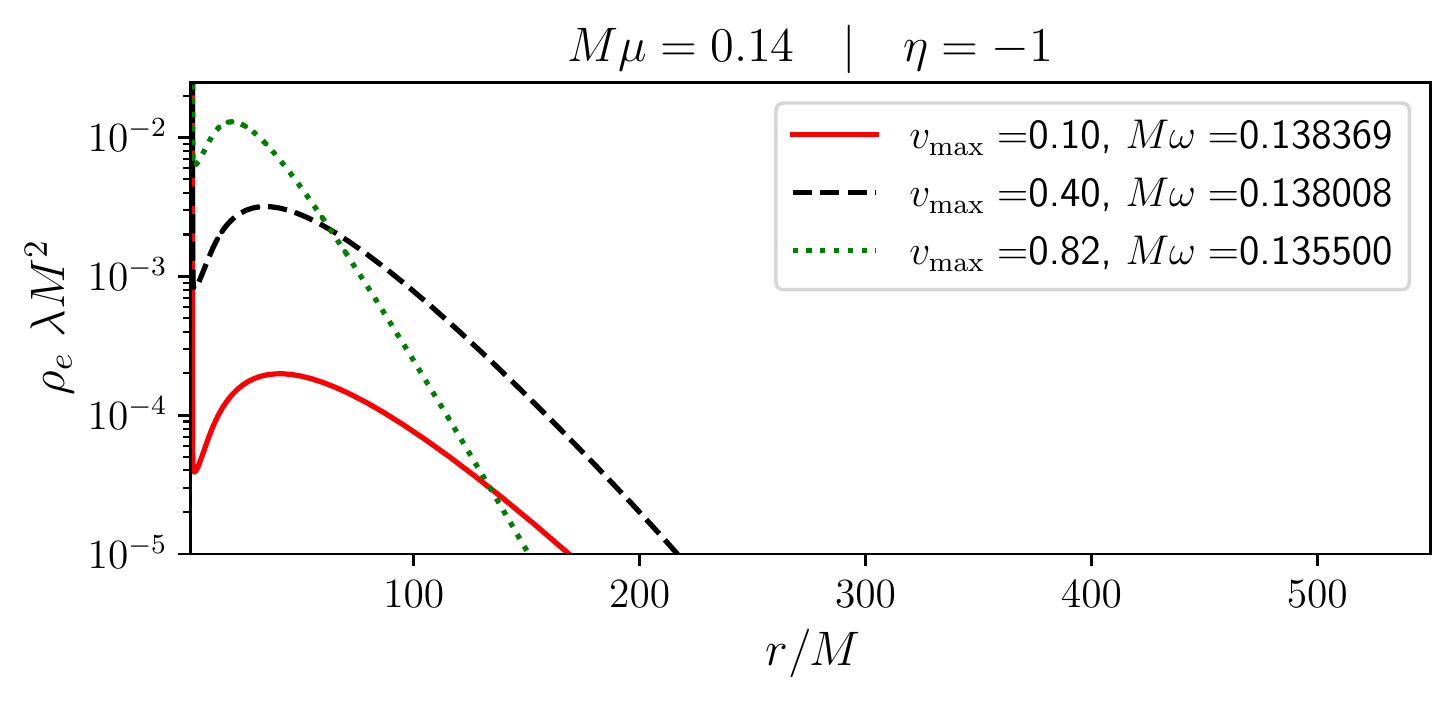}
\includegraphics[width=0.45\textwidth, angle=0]{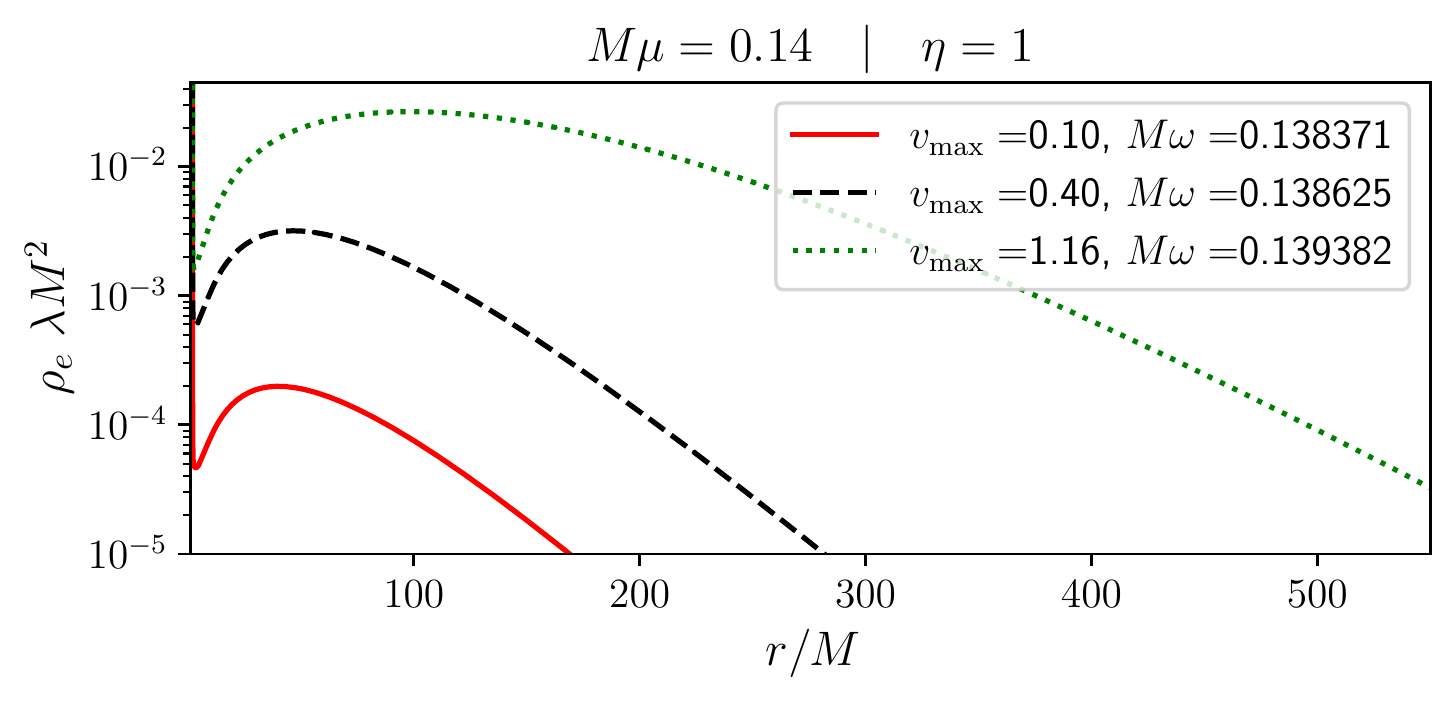}
\caption{
Radial energy density for the first resonant states for different values of $v_{\mathrm{max}}$. Left panel correspond to $\eta=-1$ whereas right panel corresponds to 
$\eta=+1$.
For $\eta=+1$
the characteristic size of the boson cloud
in the strong self-interacting regime 
is much larger than the size of the cloud in the weak self-interacting regime.
}
\label{fig:Densidades_reales}
\end{center}
\end{figure}

\begin{figure}
\begin{center}
\includegraphics[width=0.5\textwidth, angle=0]{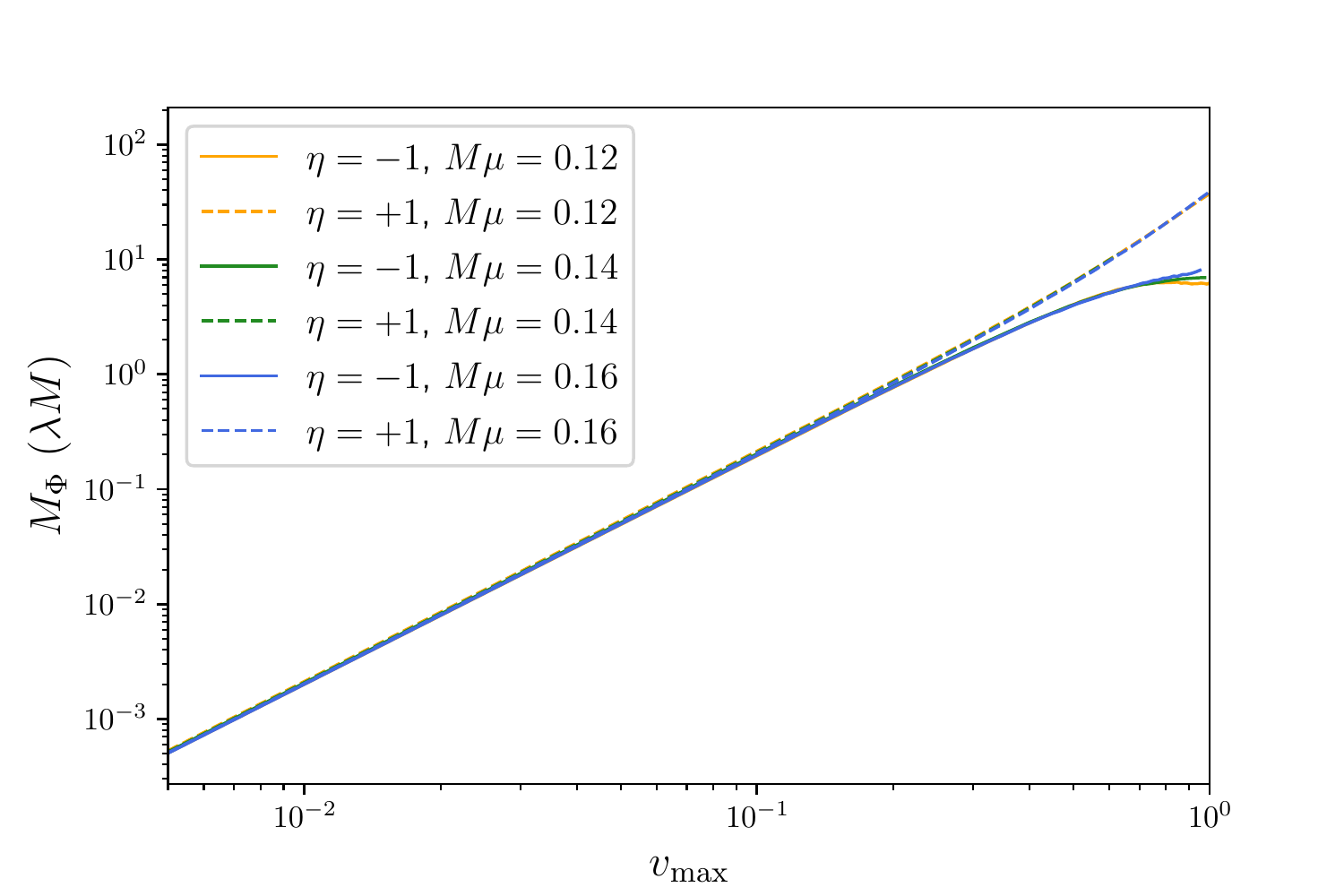}
\caption{Mass of the scalar field as defined by Eq. \eqref{eq:mass_sf}. 
In order to avoid the divergence produced by the oscillatory behaviour at the horizon,
the integration was performed from $r=2M+\epsilon$\,
to $r= r_{\rm max}$.
In practice we have taken $\epsilon$ such that $|\mathcal{U}(2M+\epsilon)|=\xi \max(|\mathcal{U}|)$, with $\xi=10^{-3}$. Other values of $\xi$ where tested with similar results.
$r_{\rm max}$ corresponds to the last point of the numerical grid.}
\label{fig:vmaxvsM_otros}
\end{center}
\end{figure}

In the following we shall focus on the fundamental resonant states as $v_{\mathrm{max}}$ increases. 
Fig.~\ref{fig:profiles_diff_eta} 
shows the scalar field profile of the fundamental resonant state for both $\eta=-1$ and $\eta=+1$. 
Each resonant state is characterized by its frequency and $v_{\rm max}$. As $v_{\rm max}$ increases the amplitude of the radial profile $v(r)$ increases for both $\eta=-1$ and $\eta=+1$.
The radial density presents the same behaviuor as is displayed in 
Fig.~\ref{fig:Densidades_reales}.
For $\eta=+1$ the effective size of the configuration increases
in the strong interacting regime.
Another important effect of the self-interaction on the distribution of the scalar field is in the mass of the configuration.
Self-interacting configurations are heavier than their non interacting partners.
In the weak self-interacting regime, the mass increases almost linearly with $v_{\rm max}$ for both
$\eta=-1$ and $\eta=+1$.
In the strong regime however, the growth slows down for $\eta=-1$. 
Fig. \ref{fig:vmaxvsM_otros} shows the mass $M_{\Phi}$ as a function of $v_{\rm max}$ for some values of the gravitational coupling $M\mu$ considering $\eta -1$ and $\eta=+1$. %

In the non self-interacting case, an effective potential of the time-independent Schrödinger-like equation \eqref{eq:KGv} (with $\lambda=0$) can be defined. In the self-interacting case, such effective potential has the form:
\begin{equation}
V_{\mathrm{eff}}(r)=N(r)\left(\frac{2M}{r^3}+\mu^2+\eta\,\lambda\,\frac{u^2(r)}{r^2}\right) \ .
\end{equation}
The effective potential interpretation may help in the characterization of the solutions (see \cite{Burt:2011pv,Barranco:2012qs} and references therein) however, for $\lambda\neq0$, $V_\mathrm{eff}$ can only be obtained \textit{a posteriori} since the knowledge of $v(r)$ (and consequently $u(r)$) is required. In any case, the effective potential may be useful to determine the resonance band formed by the parameters $M\mu$ and $M\omega$ for which the solutions have the possibility of being concentrated. The existence of the potential well in the $\lambda=0$ case is granted by the condition $M\mu<1/4$ \cite{Barranco:2012qs}.

A similar bound however, can not be found for $\lambda\neq0$.
\begin{figure}
\begin{center}
\includegraphics[width=0.45\textwidth, angle=0]{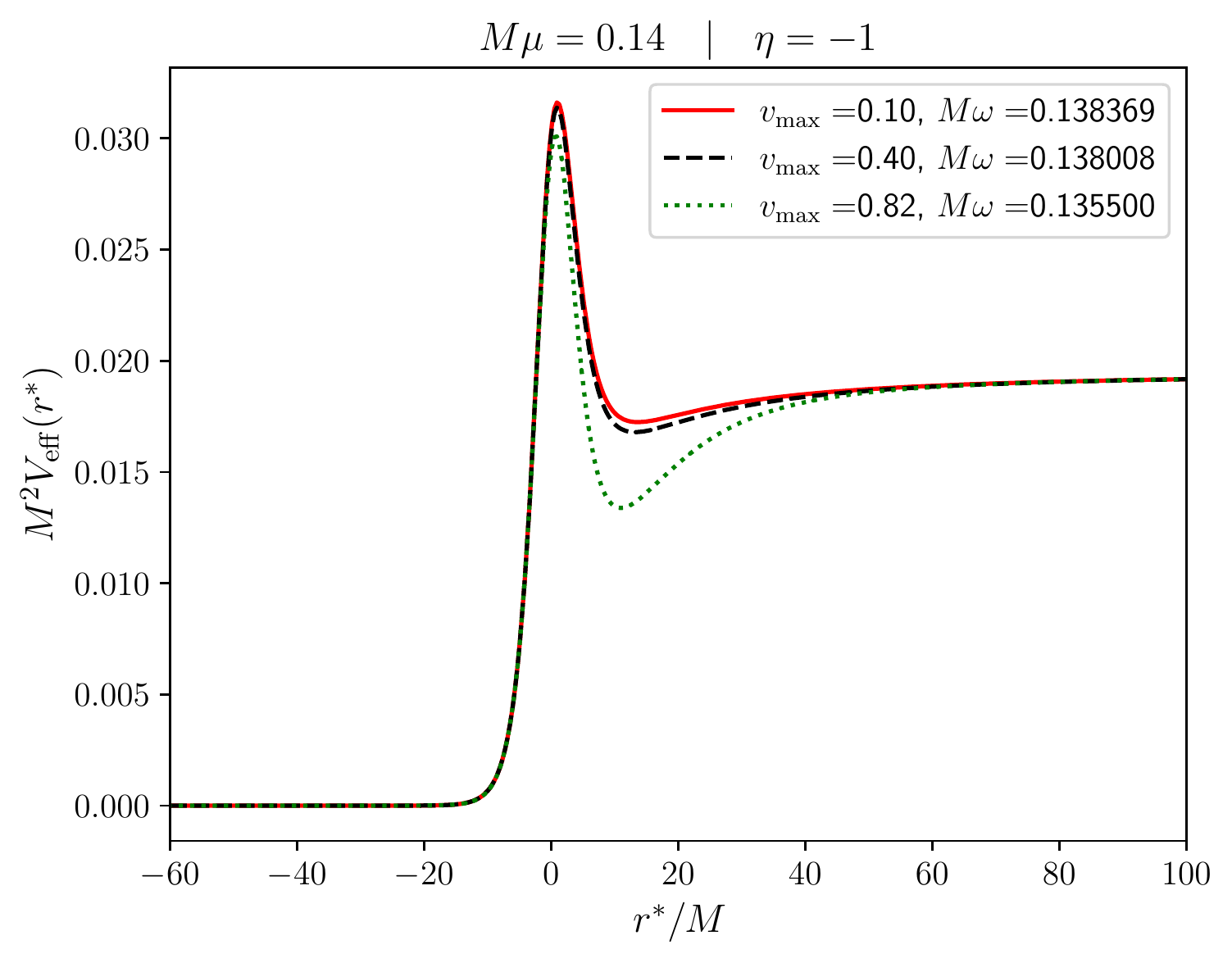}
\includegraphics[width=0.45\textwidth, angle=0]{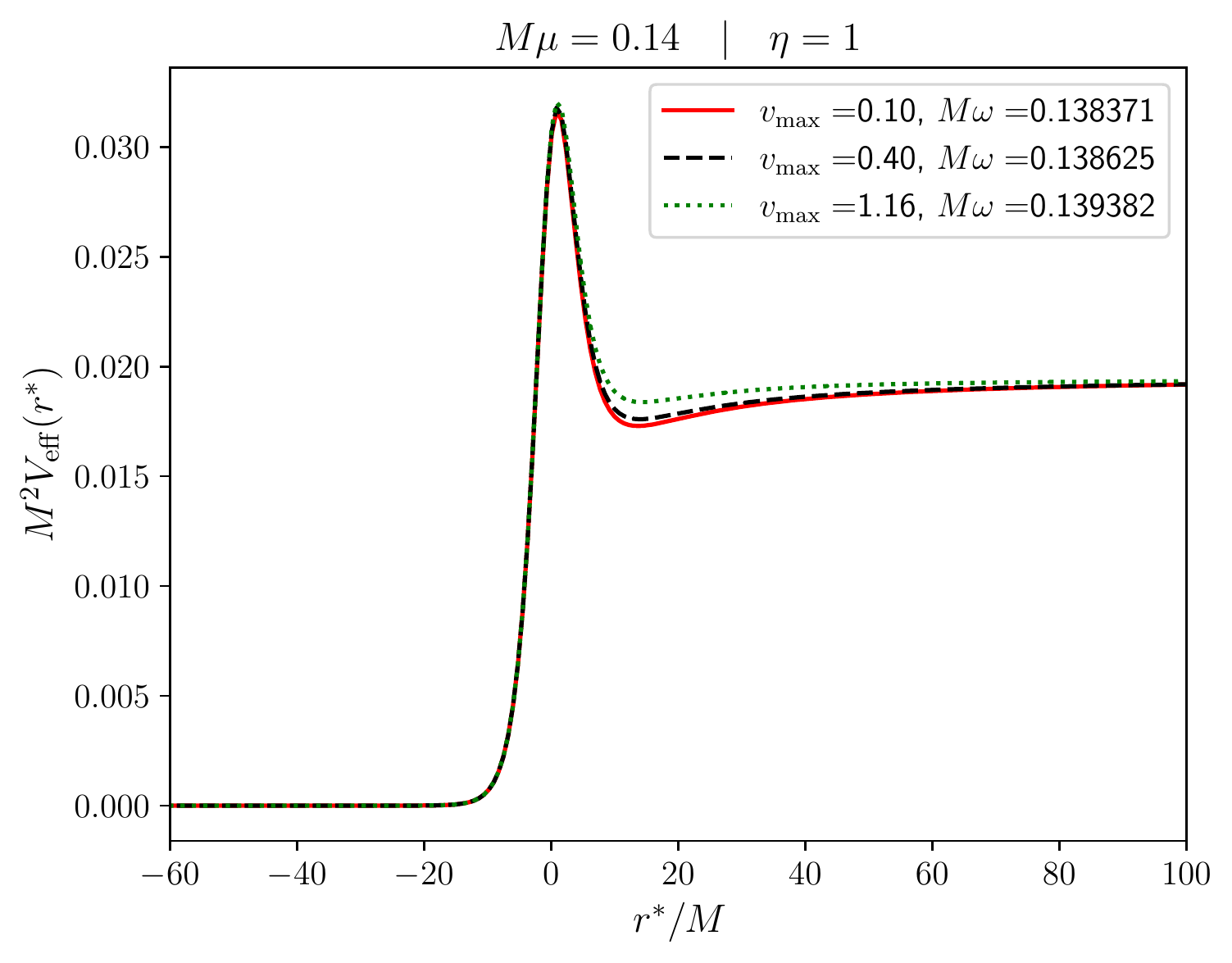}
\caption{Effective potential for the fundamental states for different values of $v_{\mathrm{max}}$ in tortoise coordinates. 
The deep of the potential well increases as $v_{\mathrm{max}}$ grows for $\eta=-1$ and decreases for $\eta=+1$. 
}
\label{fig:Veff}
\end{center}
\end{figure}

\section{Time domain description}
\label{sec:timedomain}

In this section we are interested in 
describing the behaviour in time of the radial profiles of resonant states once the harmonic time dependence \eqref{eq:decomposition} is relaxed. In particular we are interested in the life-halftime of the configuration.
To proceed, we
solve the Klein-Gordon equation in the time domain using the in-going Kerr-Schild coordinate system.
These coordinates are more convenient for numerical analysis because a 
constant time hypersurface is nonsingular and horizon penetrating. This characteristic is important in order to impose convenient boundary conditions.

The relation between the Kerr-Schild and Boyer Lindquist coordinates is given through the time transformation
\begin{equation}
\tilde t=t+(r^* - r) \, .
\end{equation}
where $r^*$ is the tortoise coordinate defined above.
In these coordinates, the Schwarzschild metric takes the form 
\begin{eqnarray}
\label{eq:ks}
ds^2 &=&-\left(1-\frac{2M}{r}\right)d\tilde t^2 + 4\frac{M}{r}drd\tilde t + \left(1+\frac{2M}{r}\right)dr^2
+ r^2 (d\theta^2+\sin^2\theta d\varphi)     \ .
 \end{eqnarray}

In terms of Arnowit Desser Misner (ADM) variables, see \cite{Corichi:1991} for an introduction, the lapse function $\alpha$, the shift vector $\beta^r$, and the induced 3-metric
can be read from Eq. \eqref{eq:ks} as
\begin{equation}
\alpha=\sqrt{\frac{r}{r+2M}}  \qquad \beta^r= \frac{2M}{2M+r} \qquad \gamma_{rr} = 1+\frac{2M}{r} \ .     
\end{equation}
The complex scalar field can be split into two real scalar fields according to 
$\Phi=\phi + i\phi_c$. However, due to the spherical symmetry of the problem at hand, both fields evolve following the same dynamical equation.
Consequently, in the following we shall refer to the dynamical properties of $\phi$.
In order to solve numerically the Klein Gordon Equation, 
\begin{eqnarray}\label{eq:kg-ks}
\nabla_{\alpha}\nabla^{\alpha} \phi=\mu^2\phi+\eta\lambda\phi^3  \ ,
\end{eqnarray}
in the background metric \eqref{eq:ks},
we introduce the auxiliary first order functions 
\begin{equation}
\psi=\partial_r \phi \qquad \pi=\alpha^2(\partial_t \phi-\beta^r\psi) \ .    
\end{equation}
where we have drop the tilde in the time coordinate. 
Equation \eqref{eq:kg-ks} is thus equivalent to the following system of evolution equations for 
$\phi$, $\pi$ and $\psi$
\begin{eqnarray}
 \partial_t \phi &=& \frac{r}{r+2M} \left(\pi+\frac{2M}{r}\psi\right) \ , \nonumber\\
 \partial_t \psi &=& \frac{2M}{2M+r} \partial_r\psi + \frac{r}{2M+r}\partial_r \pi + \frac{2M}{(2M+r)^2} \left(\pi-\psi\right) \ , \nonumber\\
 \partial_t \pi &=& \frac{1}{2M+r}\left( 2M\pi + r\psi \right) +
 \frac{2M}{(2M+r)^2}\left( \psi - \pi \right) \nonumber\\
 &&+ \frac{2}{r(2M+r)}\left( r\psi + 2M\pi\right) 
 - \phi \left(\mu^2 +\eta\lambda\phi^2 \right) \ .
\end{eqnarray}
As a diagnostic quantity we use the total energy of the scalar field that in Kerr Schild coordinates is written as 
\begin{equation}\label{eq:energy_ef}
E = 4\pi\int \rho(r) \alpha r^2 dr \ ,
\end{equation}
where
\begin{eqnarray}
\rho(r) = \frac{1}{2} 
 \left(\alpha^2\pi^2+2\pi\psi\beta^2+\gamma^{rr}\psi^2+ 
\mu^2\phi^2+\frac{1}{2}\lambda\phi^4
\right) \ .
\end{eqnarray}
The evolution equations for the radial components were solved with the 1+1 dimensional PDE solver described in \cite{Nunez:2011ej} and used in several scenarios \cite{Degollado:2013bha, Moreno:2021neu}. 
The time evolution is done via 
 the method of lines with a third order TVD Runge-Kutta algorithm. The spatial derivatives are approximated with a second order symmetric finite difference stencil. 
 A standard fourth order dissipation term was added in order to guarantee the stability of the scheme. 
 The evolution scheme is complemented imposing an outgoing wave boundary condition.

\subsection{Initial data: Resonant modes}
\begin{figure}
\begin{center}
\includegraphics[width=0.4\textwidth, angle=0]{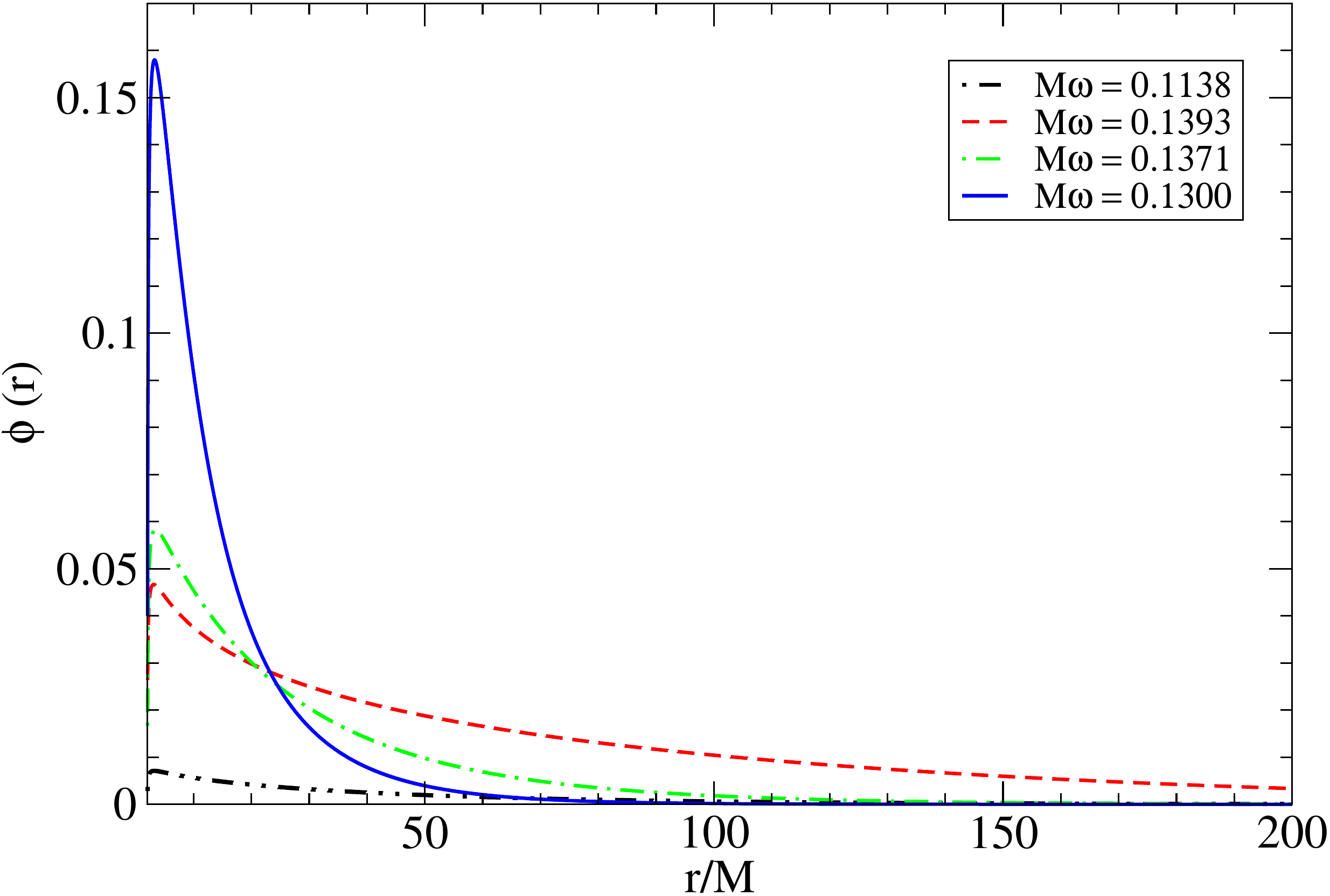}
\end{center}
\caption{
The  initial
scalar field profiles correspond to regularized resonant states for configurations with $M\mu = 0.14$. 
States with frequencies $\omega=0.13837$ and $\omega=0.1371$ are in the weakly self-interaction regime with $\eta=+1$ and $\eta=-1$, while states with 
$\omega=0.13930$ and $\omega=0.1300$ are in the strong self-interaction regime
with $\eta=+1$ and $\eta=-1$ respectively.
}
\label{fig:phi_m0.14}
\end{figure}

One may want to evolve directly a resonant state constructed in the frequency domain as described above in order to determine the time rate of decay, however, those states diverge at the horizon. In order to use the radial function of resonant states as initial data we employ a regularization procedure as described in \cite{Burt:2011pv} to mimic a quasi-resonant state and smooth its behaviour in the region close the horizon.
Such procedure allows us to set an initial data close enough to the resonant states found in previous sections. 
We have used as initial data at $t=0$ the radial profile of the field $\phi$, obtained in \ref{sec:res} with  $M\mu=0.14$.
For our analysis we consider states in both weak and strong self-interacting regimes.
Fig. \ref{fig:phi_m0.14} shows the initial radial distribution $\phi(r)$
with $\eta=+1$ and $\eta=-1$.

\begin{figure}
\begin{center}
\includegraphics[width=0.4\textwidth, angle=0]{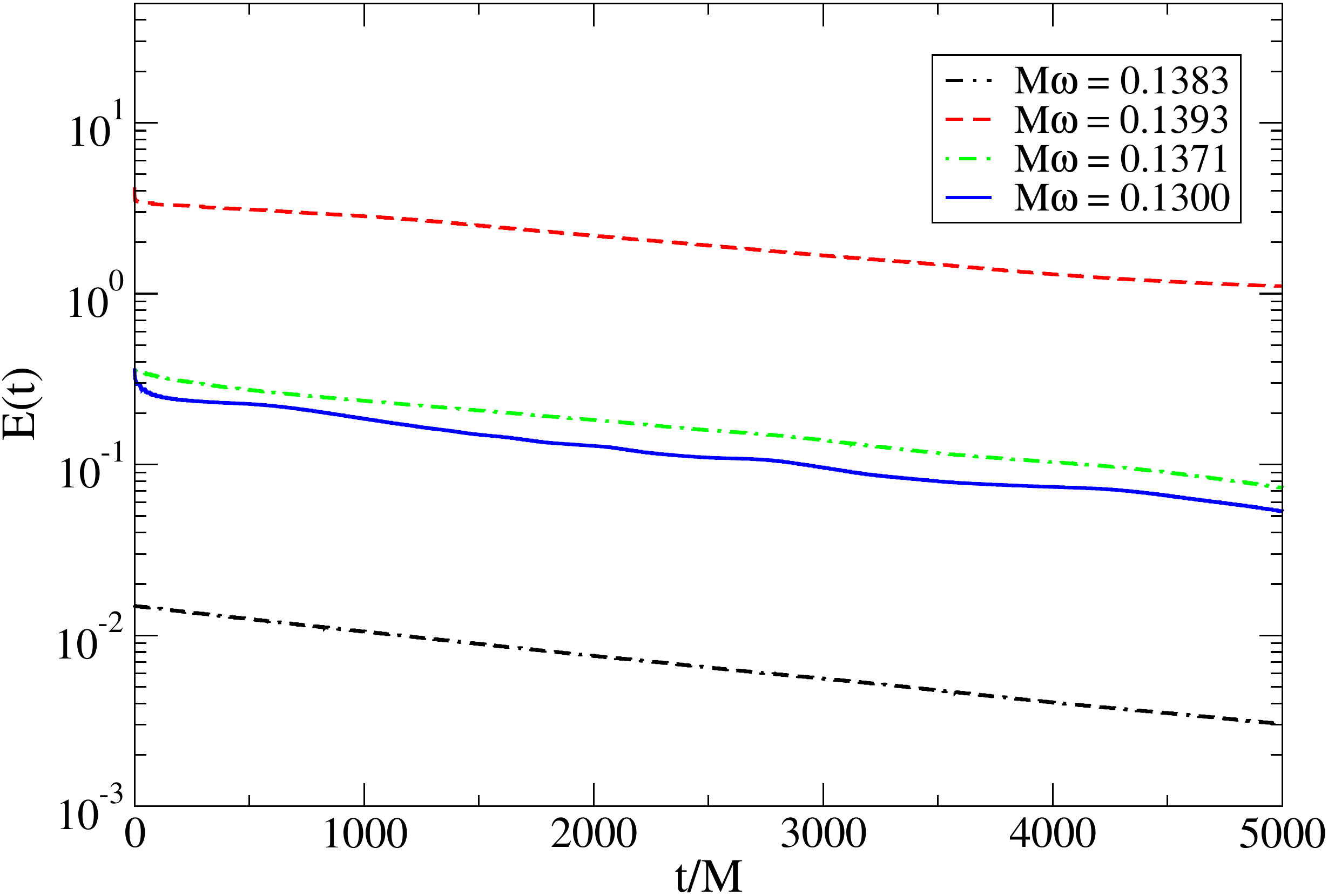}
\end{center}
\caption{Energy of the scalar field states during the evolution. After an initial fast decay, the energy decays exponentially in time. The rate of decay $s$, is shown in Table \ref{tab:resonant_l0}.}
\label{fig:ress_energy_l0}
\end{figure}

As a result of the evolution, the scalar field present an oscillating behaviour and a slow rate of decay.
By measuring the field amplitude at a fixed point $r=r_1$ one obtains a time series for the amplitude at that point. We may thus perform a Fast Fourier Transform to obtain the frequency of oscillation.
The power spectrum obtained from the time evolution, shows that the field oscillates with a frequency that
corresponds to the frequency obtained in \ref{sec:res}.
In Fig.~\ref{fig:ress_energy_l0} we also plot the total energy of the scalar configuration as a function of time $E(t)$, defined in Eq.~(\ref{eq:energy_ef}).
Fig.~\ref{fig:ress_energy_l0} shows that the decay rate of energy is exponential after an initial transient state. 
We assume that the energy has the form the form $ E(t)\sim \exp(-st)$ and 
perform a linear fit 
of $\ln(E(t)/E_0)$ with $t$
in order to find the value of the rate of $s$. The half-lifetime, of the configuration is thus given by 
$t_{1/2}/M=\ln(2)/s$.
We have further considered resonant modes with other values of the coupling $M\mu<1/4$ and compute the rate of decay of the energy.
The results are summarized in Table \ref{tab:resonant_l0}. Despite the fact the initial morphology of the scalar field state is different for configurations with
$\eta=+1$ and $\eta=-1$ the rate of decay is of the same order of magnitude for a given value of $M\mu$. 
We conclude that the effect of $\lambda$ on the rate of decay is subdominant
as compared with the effect of $M\mu$.
Notice however, that the effective size of the modes with $\eta=+1$ and large enough values of $\lambda$ is slightly larger than the non interacting case, thus if the leaking rate is of the same order of magnitude in both cases, self-interacting clouds will last longer since the initial distribution in the interacting case is larger.
Values of the boson mass $\mu$ motivated by dark matter scalar field models correspond to
$\hbar\mu \sim 10^{-24}$ eV, and for 
a black hole with mass $M\sim10^8 M_\odot$, the gravitational coupling is of the order of
$M\mu \sim 10^{-6}$. This value is slightly smaller than the ones considered in this work, such small values cannot be reached due to the limitations of the numerical integration. However if we extrapolate our findings as done in
Ref. \cite{Barranco:2013rua}  resonant modes may last 
about $10^{9}$ years.
\begin{table}
\centering
\begin{tabular}{|c |c |c |c |c ||c |} 
 \hline
$M\mu$ & $M\omega$ & $v_\mathrm{max}$ & regime &$\eta$ & $s$ \\
 \hline\hline
0.12 & 0.1192 & 0.4548 & weak & +1 & $1.16 \times 10^{-4}$\\
0.12 & 0.1196 & 1.125 & strong & +1 & $8.9\times 10^{-5}$\\
0.12 & 0.1180 & 0.6795 &  weak & $-1$ & $1.2
\times 10^{-4}$
\\
0.12 & 0.1100 & 0.9695 & strong & $-1$ &$1.9
\times 10^{-4}$
\\ 
 \hline
0.14 & 0.13837 & 0.1000 &  weak & +1 & $3.20 \times 10^{-4}$\\
0.14 & 0.13930 & 1.050 & strong & +1 & $2.64 \times 10^{-4}$\\
0.14 & 0.1371 & 0.6192 &  weak &$-1$ & $3.03 \times 10^{-4}$\\
0.14 & 0.1300 & 0.9158 &  strong &$-1$ & $3.25 \times 10^{-4}$ \\ 
 \hline
0.16 & 0.1580 & 0.4725 &  weak & +1 & $7.05\times 10^{-4}$\\
0.16 & 0.1590 & 1.102 &  strong &+1 & $4.3\times 10^{-4}$\\
0.16 & 0.1560 & 0.5564 &  weak &$-1$ &$6.9\times 10^{-4}$ \\
0.16 & 0.1500 & 0.8580 & strong & $-1$ & $5.5\times 10^{-4}$ \\ 
 \hline
0.18 & 0.1780 & 0.8232 & weak & +1 &$ 1.41 \times 10^{-3}$\\
0.18 & 0.1790 & 1.440 &  strong &+1 & $5.82 \times 10^{-4}$ \\
0.18 & 0.1750 & 0.4608 &  weak &$-1$ & 1.45 $\times 10^{-3}$ \\
0.18 & 0.1700 & 0.7803 &  strong & $-1$ & $9.6 \times 10^{-4}$ \\ 
 \hline
\end{tabular}
\caption{Parameters of the quasi-resonant states used in the evolution. The last column corresponds to the slope of the line $\ln(E(t)/E_0)=-st$ in Fig. \ref{fig:ress_energy_l0}.
}
\label{tab:resonant_l0}
\end{table}

\section{Concluding remarks}
\label{sec:conclusions}
In this paper we studied 
the role played by the self-interaction in
quasi resonant states of 
test scalar fields around a Schwarzschild black hole.
We assumed that self-interaction is mediated by a term $\sim \lambda\,\Phi^4$ and investigated the properties of the field distribution that forms in the vicinity of black holes in the case of both attractive and repulsive self-interactions. We characterized these cases by means of a parameter $\eta$, leaving the parameter $\lambda$ as a non negative quantity.
We rewrite the Klein Gordon equation in terms of a function $v(r)$, allowing us to eliminate the explicit dependence in $\lambda$ and solve it numerically. With this function one can characterize each solution with its amplitude in the near and far horizon regions. We focus our analysis on the resonant modes, which are exponentially decaying solutions at infinity concentrated well outside the event horizon. Furthermore, like in the non interacting case, resonant modes posses a definite frequency of oscillation.

The first conclusion that can be drawn 
regarding the role of self-interaction in resonant states is the fact that the values of the discrete frequencies change with respect to the non-interacting case. For $\eta=-1$ the frequency decreases and for $\eta=+1$ increases compared with the corresponding values of the frequencies without self-interaction.

We have found that in the presence of interactions the size and distribution of the scalar field changes depending on the value of $\lambda$. 
In the regime of strong self-interaction
with $\eta=+1$ the size of the resonant scalar cloud distribution increases considerably and the scalar field tend to concentrate more in a region far from the horizon as compared to the case with no self-interaction.
Even more, we have found that for large enough values of $\lambda$, resonant solutions do not exist.
Regarding the self-interaction with 
$\eta=-1$, the field concentrates closer the horizon with almost no change in size. 
In this way, we can conclude that, for $\eta=+1$ and large self-interaction
the size of the distribution is significantly larger than the
$\eta=-1$
and the non-interacting cases. 

From a classical perspective this happens because the particles of the configuration tend to concentrate outside the horizon while  gravity and attractive self-interactions tend to shrink the configuration towards the black hole.
We further investigate the life time of resonant states by means of a numerical time evolution. We found that despite the important role played by the self-interaction in the spatial distribution of the scalar field around the black hole, its role in the life time is negligible as compared to the effect of the mass term.
In the scenario described in this work 
we focus on the regime where self-interaction dominates over self gravitation effects and the parameter $\lambda$ entered in the equations as a global scale. However, when the spacetime reacts to the presence of the scalar field, the term with $\lambda$ enters explicitly on the energy density and thus a stronger effect on the properties of the whole configuration is found. A further 
study, considering the back reaction of the spacetime is under way.


\begin{acknowledgments}
This work was partially supported  by the CONACyT Network Projects No. 376127 ``Sombras, lentes y ondas gravitatorias generadas por objetos compactos astrof\'isicos", and No. 304001 ``Estudio de campos escalares con aplicaciones en cosmolog\'ia y astrof\'isica", by 
DGAPA-UNAM through grant 
IN105920 
and by the European Union's Horizon 2020 research and innovation (RISE)
program H2020-MSCA-RISE-2017 Grant
No. FunFiCO-777740. AA and VJ acknowledge  financial support from CONACyT graduate grant program. 

\end{acknowledgments}

\appendix


\section{Multi field configurations
}
\label{sec:ellg0}
In Refs.\cite{Olabarrieta:2007di} self gravitating spherically symmetric multi field configurations were considered. In these configurations an spherically symmetric tensor can be constructed assuming that the amplitudes of the constituents fields are the equal. 
The Lagrangian for $N=2\ell+1$ self-interacting complex scalar fields with a $U(N)$ symmetry is
\begin{equation}
    \mathcal{L}_{\phi}=-\sum_{i=1}^N\left(\nabla_{\mu}\Phi_i\nabla^{\mu}\Phi_i^{*}+\mu^2|\Phi_i|^2+\frac{1}{2}\lambda|\Phi_i|^2\sum_{j=1}^N |\Phi_j|^2\right),
\end{equation}
where each field is given by
\begin{align}\label{eq:decomposition_ell}
    \Phi_m(t,r,\theta,\varphi)=\frac{u(r)}{r} e^{i\omega t}{Y^{\ell m}(\theta,\varphi)\sqrt{\frac{4\pi}{2\ell+1}}}.
\end{align}
with $m=i-\ell-1$.
Using this ansatz for the fields $\Phi_m$, the resulting Klein-Gordon equation for each of the $2\ell+1$ fields is the same for all of them and is identical to Eq.~(\ref{eq:KG}) under the substitution
\begin{equation}
    \mathcal{U}(r)\rightarrow\mathcal{U}(r)+\frac{\ell(\ell+1)}{r^2}.
\end{equation}

\bibliographystyle{unsrt}
\bibliography{referencias} 

\end{document}